\documentclass[12pt,preprint]{aastex}
\usepackage{natbib}
\usepackage{amsmath,amssymb}
\usepackage{mathptmx}
\usepackage{graphicx}
\newcommand  \ergs     {\ifmmode {\rm erg\,s}^{-1} \else erg s$^{-1}$\fi}
\newcommand  \Msunyr     {\ifmmode {\rm \Msun\,yr}^{-1} \else \Msun\ yr$^{-1}$\fi}
\newcommand  \msunyr     {\ifmmode {\rm \Msun\,yr}^{-1} \else \Msun\ yr$^{-1}$\fi}
\newcommand  \cmii     {\ifmmode {\rm cm}^{-2} \else cm$^{-2}$\fi}
\newcommand  \cmiii     {\ifmmode {\rm cm}^{-3} \else cm$^{-3}$\fi}
\def\Hubble{\ifmmode {\rm km\,s}^{-1}\,{\rm Mpc}^{-1}\else km\,s$^{-1}$\,Mpc$^{-1}$\fi}
\def\Msun{\ifmmode M_{\odot} \else $M_{\odot}$\fi}
\def\msun{\ifmmode M_{\odot} \else $M_{\odot}$\fi}
\def\Lsun{\ifmmode L_{\odot} \else $L_{\odot}$\fi}
\def\Zsun{\ifmmode Z_{\odot} \else $Z_{\odot}$\fi}
\def\qo{\ifmmode q_{0} \else $q_{0}$\fi}
\def\Ho{\ifmmode H_{0} \else $H_{0}$\fi}
\def\ho{\ifmmode h_{0} \else $h_{0}$\fi}
\def\qo{\ifmmode q_{0} \else $q_{0}$\fi}
\def\ao{\ifmmode a_{0} \else $a_{0}$\fi}
\def\to{\ifmmode t_{0} \else $t_{0}$\fi}
\def\omm{\ifmmode \Omega_{{\rm M}} \else $\Omega_{{\rm M}}$\fi}
\def\omlam{\ifmmode \Omega_{\Lambda} \else $\Omega_{\Lambda}$\fi}
\def\mgii{\ifmmode {\rm Mg}{\textsc{ii}} \else Mg\,{\sc ii}\fi}
\newcommand \MgII {\ifmmode {\rm Mg}\,{\sc ii}\,\lambda2798 \else Mg\,{\sc ii}\,$\lambda2798$\fi}
\def\Hbeta{\ifmmode {\rm H}\beta \else H$\beta$\fi}

\def \L3000a{$L_{3000}$}
\def \L1450{$L_{1450}$}
\newcommand{\lbol}  {\ifmmode L_{\rm bol} \else $L_{\rm bol}$\fi}
\newcommand{\Lbol}  {\ifmmode L_{\rm bol} \else $L_{\rm bol}$\fi}
\newcommand{\lagn}  {\ifmmode L_{\rm AGN} \else $L_{\rm AGN}$\fi}
\newcommand{\LAGN}  {\ifmmode L_{\rm AGN} \else $L_{\rm AGN}$\fi}
\newcommand{\lsf}   {\ifmmode L_{\rm SF} \else $L_{\rm SF}$\fi}
\newcommand{\LSF}   {\ifmmode L_{\rm SF} \else $L_{\rm SF}$\fi}
\newcommand{\LIR}   {\ifmmode L_{\rm IR} \else $L_{\rm IR}$\fi}
\newcommand{\LHD}   {\ifmmode L_{\rm HD} \else $L_{\rm HD}$\fi}
\newcommand{\lledd} {\ifmmode L/L_{\rm Edd} \else $L/L_{\rm Edd}$\fi}
\newcommand{\Ledd} {\ifmmode L/L_{\rm Edd} \else $L/L_{\rm Edd}$\fi}
\newcommand{\fwmg}  {\ifmmode {\rm FWHM}\left(\mgii\right) \else FWHM(\mgii)\fi}
\newcommand{\CFHD}  {\ifmmode {\rm CF}_{\rm HD} \else ${\rm CF}_{\rm HD}$\fi}
\newcommand{\mbh}   {\ifmmode M_{\rm BH} \else $M_{\rm BH}$\fi}
\newcommand{\mstar}   {\ifmmode M_{*} \else $M_{*}$\fi}

\def  \mic         {$\mu$m}
\def  \MgII         {\ifmmode {\rm Mg}\,{\sc ii}\,\lambda2798
                  \else Mg\,{\sc ii}\,$\lambda2798$\fi}
\def  \mgii         {\ifmmode {\rm Mg}\,{\sc ii} \else Mg\,{\sc ii}\fi}
\def\ha{\ifmmode {\rm H}\alpha \else H$\alpha$\fi}
\def\Ha{\ifmmode {\rm H}\alpha \else H$\alpha$\fi}
\def\La{\ifmmode {\rm L}\alpha \else L$\alpha$\fi}

\def \spitzer      {{\it Spitzer}}
\def \wise {{\it WISE}}

\def \herschel {{\it Herschel}}

\def\Chisq{\ifmmode \chi^{2} \else $\chi^{2}$}
\def \zzz {\ifmmode z\simeq 4.8 \else $z \simeq 4.8$\fi} 

\def\z48{$z \simeq$4.8}
\def\z33{$z \simeq$3.3}
\def\z24{$z \simeq$2.4}

\begin{document}
% %%%%%%%%%%%%%%%%%%%%%%%%%%%%%%%%%%%%%%%%%%%%%%%%%%%%%%%%%%%%%%%%%%%%%%%%%%%%%%%%%%%%%%%%%%%%%%%

\title{Star formation and black hole growth at $z\simeq 4.8$}

\author{Hagai Netzer\altaffilmark{1},
Rivay Mor\altaffilmark{1},
Benny Trakhtenbrot\altaffilmark{2},
Ohad Shemmer\altaffilmark{3},
\& Paulina Lira\altaffilmark{4}
}

\altaffiltext{1}
{School of Physics and Astronomy and the Wise Observatory,
The Raymond and Beverly Sackler Faculty of Exact Sciences,
Tel-Aviv University, Tel-Aviv 69978, Israel}

\altaffiltext{2}
{Institute for Astronomy, Department of Physics, ETH Zurich, Wolfgang-Pauli-Strasse 27, CH-8093 Zurich, Switzerland
(Zwicky postdoctoral fellow)}

%{Department of Particle Physics and Astrophysics, The Weizmann Institute of Science, Rehovot, 76100 Israel (
%Benoziyo postdoctoral fellow) }
\altaffiltext{3}
{Department of Physics, University of North Texas, Denton, TX 76203, USA}

\altaffiltext{4}
{Departamento de Astronomia, Universidad de Chile, Camino del Observatorio 1515, Santiago, Chile}

\email{netzer@wise.tau.ac.il}

% %%%%%%%%%%%%%%%%%%%%%%%%%%%%%%%%%%%%%%%%%%%%%%%%%%%%%%%%%%%%%%%%%%%%%%%%%%%%%%%%%%%%%%%%%%%%%%%
% %%%%%%%%%%%%%%%%%%%%%%%%%%%%%%%%%%%%%%%%%%%%%%%%%%%%%%%%%%%%%%%%%%%%%%%%%%%%%%%%%%%%%%%%%%%%%%%
% %%%%%%%%%%%%%%%%%%%%%%%%%%%%%%%%%%%%%%%%%%%%%%%%%%%%%%%%%%%%%%%%%%%%%%%%%%%%%%%%%%%%%%%%%%%%%%%

\begin{abstract}
We report  \herschel/SPIRE \spitzer\ and \wise\ observations of 44 \zzz\ optically selected active 
galactic nuclei (AGNs). This flux limited sample contains the highest mass black holes (BHs) at this redshift. 
Ten of the objects were detected by \herschel\ and five show emission that is not clearly associated with the AGNs. 
The star formation (SF) luminosity (\LSF) obtained by fitting
the spectral energy distribution (SED) with standard
SF templates, taking into account AGN contribution, is
in the range $10^{46.62}-10^{47.21}$ \ergs\ corresponding to SF rates of $1090-4240$\msunyr. 
Fitting with very luminous submillimeter galaxy SEDs 
gives SF rates that are smaller by 0.05 dex when using all bands and 0.1 dex when ignoring the 250\mic\ band.
A 40K gray-body fits to only the 500\mic\ fluxes reduce \LSF\
by about a factor two.
Stacking analysis of 29 undetected sources give significant signals in all three bands.
A SF template fit indicates \LSF=$10^{46.19-46.23}$ \ergs\ depending on the assume AGN contribution.
A 40K fit to the stacked 500\mic\ flux gives \LSF=$10^{45.95}$\ergs.
The mean BH mass (\mbh) and AGN luminosity (\LAGN) of the detected sources are significantly higher 
than those of the undetected ones.
The spectral differences are seen all the way from UV to far infrared wavelengths.
The mean optical-UV spectra are similar to the ones predicted for thin  
accretion disks around BHs with the measured masses and accretion rates. 
We suggest two alternative explanations to the 
correlation of \LSF, \LAGN\ and \mbh, one involving no AGN feedback and the second a moderate feedback that affects,
but not totally quench SF in 3/4 of the sources. 
We compare our \LSF\ and \LAGN\  to lower redshift samples and show a new 
correlation between \LSF\ and \mbh.
We also examine several rather speculative ideas about the host galaxy properties
including the possibility that 
the detected sources are above the SF 
mass sequence (MS) at \zzz, perhaps in mergers, and most of the undetected sources are on the MS.

\end{abstract}


\section{Introduction}
\label{sec:introduction}

The comparison of the properties of  super-massive black holes (BHs) and their host galaxies, at various redshifts, is essential for understanding  galaxy evolution.
In particular, the star formation (SF) history in such hosts may be related to the accretion and growth of
the central BH since both are linked to the cold gas supply from the halo and the molecular gas in the disk. Such a comparison has been a very active area of
research for many years with various, occasionally conflicting results about the correlation of the luminosity of the active galactic nucleus (AGN)  (\LAGN) and the
SF rate (SFR), or SF luminosity (\LSF) is such systems \cite[e.g.][and references therein]{Rosario2012,mullaney2012}.

 The launch of \herschel\ was an important mile-stone in this area. This mission allowed deeper and more 
accurate measurements of \LSF\ through sensitive, far infrared (FIR) observations. It is becoming clear
that most AGNs, at all redshifts, reside in SF galaxies \citep{Silverman2008,Santini2012,Mullaney2012a,Mainieri2011}
and there is no evidence for different properties of the host, whether it contains an active or dormant
BH, provided the stellar mass (\mstar) is the same  \citep{Rosario2013}.
Several \herschel-based studies, e.g. \cite{Shao2010},  \cite{Hatziminaoglou2010}, \cite{Rosario2012}, \cite{Harrison2012} and \cite{Page2012}
show the complex dependence of \LSF\ on \LAGN\ at all redshifts up to $z \sim 2.5$. Some of the studies \citep{Shao2010,Rosario2012} demonstrate that  the \LSF-\LAGN\ plane can be divided
into two regimes with very different behaviors. One is the ``SF dominated'' regime where \LSF$>$\LAGN. Here the two properties are not correlated and \LSF\  exceeds
\LAGN\ by a redshift dependent factor.
In the second ``AGN dominated'' regime, where \LAGN$>$\LSF, the sources seem to confluence around a power-law line which is given roughly by
 $\lsf\simeq10^{43}(\lagn/10^{43} \ergs)^{0.7}$. A very similar 
relationship was suggested in earlier works 
\cite[e.g.,][]{Netzer2007b,Lutz2008,Netzer2009a}. The exact distribution in the \LSF-\LAGN\ plane is still to be verified, because of the various selection
effects entering the selection of FIR-faint high redshift sources \citep[e.g.][]{Page2012,Harrison2012}, and is still to be explained by theoretical models. 

The situation at $2<z<4$ is more complex because almost all of the \herschel\ observations are affected by
confusion noise. This is the reason why, so far, there are very few systematic FIR studies at such redshifts. At even   
higher redshifts there are very few known SF-dominated sources
although a handful of sources with \lsf$\simeq$\lagn\ have been found mainly by sub-mm observations
\cite[e.g.,][]{Isaak2002,Priddey2003,Wu2009,Leipski2010,Omont2013,Wang2011}. 
It is therefore important to study well defined samples, selected by their optical properties, using \herschel.
This is a major aim of the present work.

Galaxy evolution scenarios suggest two modes of SF.
The steadier process (hereafter ``secular evolution'') is common in isolated disk galaxies and can reach SFRs of $\sim 400$ \Msunyr\ for the most massive galaxies
at high redshift. SF due to starburst is generally less common. The SFR in such cases can exceed $\sim$1000 \Msunyr\
and the starburst activity 
 is associated, in many cases, with mergers between galaxies \citep{Wuyts2011a,Wuyts2011b}.
The highest SFR cases are thought to be associated with mergers of similar mass, massive 
gas-rich galaxies \cite[see e.g.,][]{Rodighiero2011}.
Both secular evolution and starburst through mergers can result in cold gas inflow into the center of the system which can trigger AGN activity
\cite[][]{DiMatteo2005,Guyon2006,Sijacki2011,Valiante2011}.

A somewhat different way to address SFR, especially at high redshift, is to consider feeding of isolated galaxies by ``cold streams'' from the halo.
Hydrodynamical simulations of such processes, like those published recently by \cite{Khandai2012}, show a strong correlation between 
SFR and BH accretion rate at
$z \sim 5$. The SFR in such systems can reach and even exceed 1000\msunyr.

It is not yet clear which of the gas supply mechanisms is more important and under what conditions. 
A recent semi-analytic model by \cite{Neistein2013} shows that many observed
correlations at $z<2.5$, including the general behavior in the \LSF-\LAGN\ plane, can be reproduced by a model where AGN activity is triggered {\it solely} by mergers.
Other numerical simulations and semi-analytic models suggest that in merger events, the fastest SMBH growth phase succeeds the
peak of SF activity by several hundred Myr. A related suggestion is that the earlier stages of a major merger take place when the BH is obscured \cite[e.g.,][]{Hopkins2006a,DiMatteo2008}.
Some support for this idea comes from the fact that sub-mm galaxies (SMGs) with very high SFR, often exhibit little or no AGN activity. However, there is no complete census of
the AGN population at high redshift  to support this claim. 
Finally, AGN feedback, that can quench  SF and BH accretion through  fast winds and  photoionization by the
intense AGN radiation \cite[e.g.,][]{DiMatteo2005, Springel2005, Sijacki2007}, can also contribute to the relationship between \LSF\ and \LAGN\ in systems
hosting high luminosity AGNs.

It this paper we discuss the UV, optical and IR properties of an optically selected flux limited sample of AGNs
at \zzz. The basic physical properties of the sources are given
in \cite[][hereafter T11]{Trakhtenbrot2011} and preliminary 
\herschel\ results for 25 of the sources where published in \cite[][hereafter M12]{Mor2012}. 
Here we expand the FIR study to all 44 sources and include also new \spitzer\ and \wise\ observations.
Section~\ref{sec:observations} introduces all the observations and the flux and luminosity measurements and
\S~\ref{sec:discussion} discusses various central issues like the correlations between \LSF, \LAGN, and BH mass (\mbh). We also address the properties of the UV-FIR spectral
energy distribution (SED) of the sources
and suggest various evolutionary scenarios that connect the accumulation of BH and stellar mass through cosmic time.
Throughout this paper we assume $\Ho=70$ \Hubble, $\omm = 0.3$ and $\omlam = 0.7$.

%%%%%%%%%%%% Herschel SPIRE reduction and photometry

\section{Observations Reduction and Basic Analysis}
\label{sec:observations}

\subsection{\herschel\ observations}

The AGN sample described in this work includes 44 high luminosity AGNs at $z \simeq 4.8$. This is a flux limited
sample selected from the Sloan Digital Sky Survey (SDSS, York et al 2000)
 to include the most luminous AGNs in a narrow redshift range 4.66--4.87.
The sample was described in T11 where details of the H and K-band spectroscopy are given. BH mass measurements are available for 40 of the sources using the 3000\AA\ continuum
luminosity and the \MgII\ line width (see T11 for more details).
The additional 4 objects were observed by us but resulted in poor quality \MgII\
observations and hence no reliable mass estimates. The observations 
allow good measurements of \mbh,  \Ledd\ and an estimate of the duty cycle of this population.

The \herschel\ observations cover all original 44 sources including the 4 objects with no reliable BH mass determination.
Data reduction and analysis utilized standard \herschel\ tools and
was explained in detail in M12. Here we summarize the more important details for clarity.

Details of the \herschel\ observations are given in Table~\ref{tab:data}.
All sources have been observed with the SPIRE instrument \cite[]{Griffin2010}  providing images at 250, 350, and 500 \mic, corresponding to the 
rest-frame FIR wavelength range of 43--86 \mic. 
The observations were made in the small-map mode. 
The data reduction process starts with the level 0.5 product of the SPIRE pipeline and uses standard tools provided by the \herschel\ Science Centre (HSC) 
via the HIPE software \cite[][version 9.0]{Ott2010}, 
and version 9.1 of the calibration files.

\begin{table}
\begin{center}
\tiny
\caption{Observed Properties}\label{tab:data}
\begin{tabular}{llccccccccl} 

\hline \hline
                   &	   & \multicolumn{4}{l}{\spitzer\ observations$^{\rm b}$} &  \multicolumn{5}{l}{\herschel\ observations$^{\rm c}$} \\
Object             &z$^{\rm a}$&Spitzer ID&$F_\nu$(3.6\mic)& $F_\nu$(4.5\mic)& $N_{\rm near}^{\rm d}$& SPIRE  &$F_\nu$(250\mic)&$F_\nu$(350\mic)& $F_\nu$(500\mic) & FIR \\
(SDSS J)           &          &  (AOR\#) & ($\mu$Jy)	     & ($\mu$Jy)       &                       & Obs. ID&(mJy)      &(mJy) 	       & (mJy) 		   & status$^{\rm e}$\\
\hline
000749.16+004119.6 & 4.786    & 42390016 &  40.95$\pm$1.31 &  37.35$\pm$0.76 & 1& 1342212418 &  0.0$\pm$5.9 &  0.0$\pm$6.3 &  0.0$\pm$6.8 & o\\ %02
003525.28+004002.8 & 4.759    & 42390272 & 119.15$\pm$1.33 &  89.76$\pm$0.79 &  & 1342213190 &  0.0$\pm$5.8 &  0.0$\pm$6.3 &  0.0$\pm$6.8 & -\\ %03
021043.16-001818.4 & 4.713    & 34925056 &  80.69$\pm$2.78 &  66.12$\pm$2.22 &  & 1342237543 &  0.0$\pm$5.8 &  0.0$\pm$6.3 &  0.0$\pm$6.8 & -\\ %40
033119.66-074143.1 & 4.729    & 42390528 & 148.83$\pm$1.33 & 113.81$\pm$0.81 & 1& 1342214563 & 26.4$\pm$6.1 & 24.6$\pm$6.6 & 22.1$\pm$7.0 & +\\ %05
075907.57+180054.7 & 4.804    & 42390784 & 167.70$\pm$1.34 & 124.09$\pm$0.81 &  & 1342229465 &  0.0$\pm$5.8 &  0.0$\pm$6.3 &  0.0$\pm$6.8 & -\\ %18
080023.01+305101.1 & 4.677    & 42391040 & 179.71$\pm$1.35 & 138.80$\pm$0.82 &  & 1342229475 &  0.0$\pm$5.8 & 25.9$\pm$6.6 & 35.5$\pm$7.3 & -\\ %19
080715.11+132805.1 & 4.885    & 42391296 &  85.66$\pm$1.31 &  72.70$\pm$0.77 & 2& 1342230781 & 12.8$\pm$5.9 & 19.3$\pm$6.5 & 13.7$\pm$6.9 & +\\ %26
083920.53+352459.3 & 4.795    & 42391552 &  49.97$\pm$1.30 &  37.48$\pm$0.76 &  & 1342230757 &  0.0$\pm$5.8 &  0.0$\pm$6.3 &  0.0$\pm$6.8 & -\\ %24
085707.94+321031.9 & 4.801    & 42391808 & 194.19$\pm$1.35 & 149.68$\pm$0.81 &  & 1342230758 &  0.0$\pm$5.8 &  0.0$\pm$6.3 &  0.0$\pm$6.8 & -\\ %25
092303.53+024739.5 & 4.659    & 42392064 &  59.43$\pm$1.32 &  48.52$\pm$0.75 &  & 1342245156 &  0.0$\pm$5.8 &  0.0$\pm$6.3 &  0.0$\pm$6.8 & -\\ %42
093508.49+080114.5 & 4.671    & 42392320 & 101.10$\pm$1.32 &  85.28$\pm$0.79 &  & 1342245560 &  0.0$\pm$5.8 &  0.0$\pm$6.3 &  0.0$\pm$6.9 & -\\ %43
093523.31+411518.5 & 4.802    & 42392576 &  76.47$\pm$1.30 &  63.74$\pm$0.75 &  & 1342230746 &  0.0$\pm$5.8 &  0.0$\pm$6.3 &  0.0$\pm$6.8 & -\\ %23
094409.52+100656.6 & 4.771    & 42392832 & 168.76$\pm$1.35 & 128.43$\pm$0.80 &  & 1342246601 &  0.0$\pm$5.8 &  0.0$\pm$6.3 &  0.0$\pm$6.8 & -\\ %44
101759.63+032739.9 & 4.943    & 42393088 &  55.95$\pm$1.32 &  43.00$\pm$0.87 &  & 1342222673 &  0.0$\pm$5.8 &  0.0$\pm$6.4 &  0.0$\pm$6.8 & -\\ %10
105919.22+023428.7 & 4.789    & 42393344 &  70.43$\pm$1.33 &  56.78$\pm$0.82 &  & 1342222891 &  0.0$\pm$5.9 & 20.6$\pm$6.5 &  0.0$\pm$7.0 & o\\ %13
110045.23+112239.1 & 4.707$^*$&          &                 &                 &  & 1342222887 &  0.0$\pm$5.8 &  0.0$\pm$6.3 &  0.0$\pm$6.8 & -\\ %11
111358.32+025333.6 & 4.870    & 42393600 &  86.44$\pm$1.34 &  72.52$\pm$0.85 & 1& 1342222890 &  0.0$\pm$5.8 &  0.0$\pm$6.3 &  0.0$\pm$6.9 & o\\ %12
114448.54+055709.8 & 4.790    & 42393856 &  42.61$\pm$1.30 &  31.84$\pm$0.78 &  & 1342234877 &  0.0$\pm$5.8 &  0.0$\pm$6.3 &  0.0$\pm$6.8 & -\\ %30
115158.25+030341.7 & 4.687    & 42394112 &  22.79$\pm$1.29 &  18.54$\pm$0.76 &  & 1342234878 &  0.0$\pm$5.8 &  0.0$\pm$6.4 &  0.0$\pm$6.8 & -\\ %31
120256.43+072038.9 & 4.810    & 42394368 &  88.40$\pm$1.35 &  74.90$\pm$0.91 &  & 1342234895 &  0.0$\pm$5.8 &  0.0$\pm$6.4 &  0.0$\pm$6.8 & -\\ %33
123503.03-000331.7 & 4.700    & 42394624 &  69.66$\pm$1.34 &  62.81$\pm$0.95 &  & 1342234884 &  0.0$\pm$5.9 &  0.0$\pm$6.4 &  0.0$\pm$6.8 & -\\ %32
130619.38+023658.9 & 4.860    & 42394880 & 239.70$\pm$1.39 & 207.72$\pm$0.98 &  & 1342224983 &  0.0$\pm$5.9 &  0.0$\pm$6.4 &  0.0$\pm$6.8 & -\\ %14
131737.27+110533.0 & 4.744    & 42395136 & 116.79$\pm$1.34 &  89.78$\pm$0.89 &  & 1342224984 &  0.0$\pm$5.8 &  0.0$\pm$6.3 &  0.0$\pm$6.8 & o\\ %15
132110.81+003821.7 & 4.726    & 42395392 &  69.19$\pm$1.34 &  61.81$\pm$0.93 &  & 1342224986 &  0.0$\pm$5.8 &  0.0$\pm$6.3 &  0.0$\pm$6.8 & -\\ %17
132853.66-022441.6 & 4.658    & 42395648 &  82.86$\pm$1.32 &  64.65$\pm$0.87 &  & 1342234799 &  0.0$\pm$5.8 &  0.0$\pm$6.3 &  0.0$\pm$6.8 & -\\ %29
133125.56+025535.5 & 4.762    & 42395904 &  30.33$\pm$1.32 &  28.64$\pm$0.88 &  & 1342224985 & 23.0$\pm$6.1 & 38.5$\pm$7.0 & 36.8$\pm$7.4 & o\\ %16
134134.19+014157.7 & 4.689    & 42396160 & 159.69$\pm$1.36 & 132.01$\pm$0.95 & 2& 1342234797 & 41.7$\pm$6.6 & 47.4$\pm$7.3 & 42.7$\pm$7.5 & +\\ %28
134546.96-015940.3 & 4.728    & 42396416 &  63.72$\pm$1.33 &  48.23$\pm$0.91 &  & 1342236167 &  0.0$\pm$5.8 &  0.0$\pm$6.3 &  0.0$\pm$6.8 & -\\ %39
140404.63+031403.9 & 4.903    & 42396672 &  97.72$\pm$1.34 &  86.83$\pm$0.90 &  & 1342236163 & 25.8$\pm$6.1 & 31.1$\pm$6.7 & 22.9$\pm$7.0 & +\\ %38
143352.21+022713.9 & 4.722    & 42396928 & 302.75$\pm$1.39 & 295.18$\pm$0.93 &  & 1342236160 & 28.1$\pm$6.2 & 25.7$\pm$6.6 & 12.7$\pm$6.9 & +\\ %37
143629.94+063508.0 & 4.817    & 42397184 &  71.39$\pm$1.31 &  54.55$\pm$0.81 &  & 1342236156 &  0.0$\pm$5.8 &  0.0$\pm$6.3 &  0.0$\pm$6.8 & -\\ %35
144352.94+060533.1 & 4.884    & 42397440 &  48.02$\pm$1.29 &  34.74$\pm$0.78 &  & 1342236159 &  0.0$\pm$5.8 &  0.0$\pm$6.3 &  0.0$\pm$6.8 & -\\ %36
144734.09+102513.1 & 4.679    &          &                 &                 &  & 1342236153 &  0.0$\pm$5.8 &  0.0$\pm$6.3 &  0.0$\pm$6.8 & -\\ %34
151155.98+040802.9 & 4.670    & 42397696 &  92.58$\pm$1.32 &  87.45$\pm$0.80 & 2& 1342238320 & 26.6$\pm$6.1 & 38.0$\pm$6.9 & 28.9$\pm$7.2 & +\\ %41
161622.10+050127.7 & 4.869    & 42397952 & 151.59$\pm$1.33 & 122.64$\pm$0.83 & 1& 1342229564 & 48.0$\pm$6.9 & 49.7$\pm$7.4 & 47.4$\pm$7.7 & +\\ %21
161931.58+123844.4 & 4.806$^*$&          &                 &                 &  & 1342229563 & 47.7$\pm$6.8 & 38.5$\pm$7.0 & 21.6$\pm$7.0 & +\\ %20
165436.85+222733.7 & 4.717    & 42398208 & 248.00$\pm$1.35 & 204.63$\pm$0.78 & 1& 1342229582 & 21.9$\pm$6.0 & 15.0$\pm$6.4 &  5.9$\pm$6.8 & +\\ %22
205724.14-003018.7 & 4.680    &          &                 &                 &  & 1342218986 &  0.0$\pm$5.9 &  0.0$\pm$6.3 &  0.0$\pm$6.8 & -\\ %06
220008.66+001744.9 & 4.804    & 42398464 & 135.76$\pm$1.34 & 101.04$\pm$0.79 &  & 1342233332 &  0.0$\pm$5.8 &  0.0$\pm$6.3 &  0.0$\pm$6.8 & -\\ %27
221705.71-001307.7 & 4.676    & 34924288 &  57.33$\pm$2.32 &  51.24$\pm$1.64 &  & 1342220873 &  0.0$\pm$5.8 &  0.0$\pm$6.3 &  0.0$\pm$6.8 & -\\ %09
222050.80+001959.1 & 4.716$^*$& 34924544 &  53.97$\pm$2.32 &  48.54$\pm$1.65 &  & 1342220872 &  0.0$\pm$5.8 &  0.0$\pm$6.3 &  0.0$\pm$6.8 & -\\ %08
222509.19-001406.9 & 4.890    & 34925312 & 156.34$\pm$2.45 & 125.59$\pm$1.75 &  & 1342220530 & 24.7$\pm$6.1 & 30.3$\pm$6.7 & 22.2$\pm$7.0 & +\\ %07
224453.06+134631.6 & 4.656    & 42398720 &  48.04$\pm$1.29 &  48.86$\pm$0.73 &  & 1342211361 &  0.0$\pm$5.8 &  0.0$\pm$6.4 &  0.0$\pm$6.8 & -\\ %01
235152.80+160048.9 & 4.694$^*$&          &                 &                 &  & 1342213200 &  0.0$\pm$5.8 &  0.0$\pm$6.3 &  0.0$\pm$6.8 & -\\ %04
\hline
\end{tabular} 
\end{center}
$^{\rm a}$ determined by T11 from the \mgii\ line, except for objects with asterisks, for which redshifts are taken from the SDSS DR7 QSO catalog.\\
$^{\rm b}$ flux densities are measured through an aperture of 3.6 arcsec.\\
$^{\rm c}$ flux densities are measured through apertures centered on the SDSS source centers. For sources with $N_{\rm near}>0$, the flux measurements take into account the neighboring sources (see \S2.5).\\
$^{\rm d}$ number of nearby \spitzer\ sources (within $4\times$PSF).\\
$^{\rm e}$ classification of \herschel\ data: ``+'' - detected; ``-'' - nondetected; ``o'' - offset.
\normalsize
\end{table}

\subsection{\herschel\ photometry}

Nine out of 44 sources, all with measured, \mbh,  are clearly detected in all three SPIRE bands.
We follow the guidelines of the HSC and apply a peak fitting method to all images of these sources (M12).
in order to measure the FIR flux. 
Due to the low resolution of the SPIRE images, and the pointing accuracy of \herschel\
 (about $5"$) the determination of the 
the exact location of the source on the image is somewhat uncertain. We therefore apply our photometry twice with the two 
dimensional Gaussian centered at two different locations. 
The first is determined by the pointing of the \herschel\ telescope and the second is the center of the SDSS image. 
Note that even with perfect pointing, the two do not necessarily
agree since the first represents the peak of SF activity and the second the location of the active BH. At $z=4.8$, $1"$ 
corresponds to 6.4 kpc and the difference
in location between the two is likely a tiny fraction of the point spread function (PSF) of the instrument. 
The flux measurements for the two locations agree very well with a difference which is, at most, 5\%.
Table~\ref{tab:data} lists fluxes measured assuming the object location is determined by the pointing of the telescope\footnote{The other fluxes are available from the authors upon request}.
350\mic\ images of several detected sources are given in M12. 

There are three known sources of uncertainty related to SPIRE photometry
(see the SPIRE observers` manual\footnote{http://herschel.esac.esa.int/Docs/SPIRE/html/spire\_om.html}).
 The first is related to the fitting procedure and includes both instrumental and confusion noise. The integration times in our program  
(either 222 or 296 seconds) are long enough to minimize the instrumental noise
and obtain images which are dominated by the extragalactic confusion noise estimated 
to be 5.8, 6.3 and 6.8 mJy/beam at 250, 350 and 500 \mic, respectively \cite[]{Nguyen2010}. 
The second uncertainty is related to the pixelization correction of the images. This introduces an additional uncertainty of about 2--3\% 
to the flux density. 
The third uncertainty is associated with the calibration process and is about 7\% of the flux density. 
These uncertainties are added in quadrature and listed together with the measured fluxes in Table~\ref{tab:data}.

Six of the sources, all with measured \mbh, show clear signs of emission within the instrument PSF in all bands. However, the overall
shape is very different from a point source and the peak emission is offset by several pixels from the expected SDSS location
of the AGN which corresponds to more than 100 kpc at $z=4.8$. We refer to these objects as ``offset'' sources. As shown below,
 our prior-based
analysis suggests that one of these sources includes a real detection in the  location of the AGN. 

\subsection{\spitzer\ Observations and photometry}

We obtained additional observations of most of the sources in our sample using the \spitzer\ Infrared Array Camera 
(IRAC; Fazio et al. 2004) in Cycle 8 (Program ID 80093, PI O. Shemmer). 
In total 35  of the sources were observed by \spitzer/IRAC in the 3.6 and 4.5 \mic\ bands. Details of the observations are given in Table~\ref{tab:data}. 
The net integration time for each observation is 360 seconds. All sources are clearly detected in both \spitzer\ bands.
%(total AOR time of 1118 seconds).
In addition, four sources in our sample have archival IRAC images (Program ID 60139; PI G. Richards). 
These images were reduced and analyzed in a way identical to the images obtained by us. It total, we have 39 
\spitzer\ detections in both IRAC bands. Nine of these sources correspond to \herschel-detected AGNs with measured \mbh\
and 24 to \herschel-undetected AGNs with measured \mbh..

The reduction of the IRAC observations utilized standard \spitzer\ 
routines and will not be described in detail. It is based on the
version S18.5.0 of the 
\spitzer\ Science Center (SSC) pipeline and uses the specific software package {\sc mopex} (version 18.5.4; Makovoz \& Marleau 2005).
 We have measured the flux in several available apertures and used the correction factors
provided in the IRAC
Handbook\footnote{http://irsa.ipac.caltech.edu/data/SPITZER/docs/irac/iracinstrumenthandbook/home/}. 
We chose to use the recommended aperture of $3.6"$.
Array-location-dependent corrections were applied to the IRAC data but no 
color or pixel-phase corrections were used as these were found to have a minute ($<1$\%) effect on the total flux. 
Table~\ref{tab:data} presents the results of our \spitzer\ photometry.  The statistical flux uncertainties for all mosaics are 
smaller than the 5\% calibration accuracy of IRAC. However, 
the differences between the 3.6 and 6 \arcsec\ fluxes, including the aperture correction
factors, are larger, of order 10\%.  In the table we quote this 10\%\ uncertainty for all our measured fluxes. 

The IRAC observations provide data on the AGN continuum at effective rest-frame wavelengths of about 6120 and 7760\AA.
The strong \ha\ line flux is included
in the 3.6\mic\ band and must be taken into account when considering the SED of the AGNs. 
Based on observations of strong-line type-I AGNs \citep{Stern2012}, we estimate that the line contribution to the total flux in
this band  is $\sim 25$\%.
Below we refer to individual \spitzer\ observations\ as well as to two sub-groups corresponding to the
\herschel\ sub-groups of detected and
undetected sources. 

\subsection{\wise\ Observations and photometry}

We have used publicly available data from the Wide-Field Infrared Survey \cite[\wise][]{Wright2010}) to complement our \spitzer\ data.
The all-sky point source catalog was queried for entries that are located within 5\arcsec\ of the SDSS positions of the $z\simeq4.8$ sources.  
This query provided detections (i.e. sources with $>5\sigma$ in at least the 3.35 $\mu{\rm m}$, or ``W1'', band) for 35 sources. 
Of these, 31, 11 and 4 sources had significant detections ($>3\sigma$) in the 4.6, 11.56 and 22.1 $\mu{\rm m}$ (``W2'', ``W3'' and ``W4'') bands, respectively.
The cataloged magnitudes were converted to flux densities using the conversion factors tabulated in \cite{Wright2010} (table 1). 
The specific choice of assumed SEDs has a negligible effect (a few percent) on the derived flux densities. 
Symmetric uncertainties on the flux densities were derived directly from the S/N ratios (SNRs) listed in the catalog. 
These flux densities and uncertainties are listed in Table~\ref{tab:data}. 
We note that the uncertainties, and thus the level of significance,  differ significantly across the sample. This is due to the heterogeneous 
nature of the \wise\ coverage. Thus, the data do not provide a complete mid-IR survey of the \zzz\ sample.

The two shorter-wavelength \wise\ bands probe essentially the same wavelength range as the \spitzer\ observations. Indeed, the observed fluxes are
very similar in all cases with median differences of less than 0.07 and 0.002 dex, and standard deviations of  about 0.12 and 0.11 dex, for the first and second bands, 
respectively (24 and 20 overlapping objects). These bands probe the AGN-dominated continuum and some of the  
differences may be due to source variability.

The very small number of sources detected in the 22\mic\ (``W4’’) band prevents us from making use of this information.
Hence, we combine the \wise\ data in the two short-wavelength bands with the \spitzer\ data and conduct a separate analysis of 
the 11.6\mic\ (``W3’’) band as discussed below.

\subsection{\herschel\ Photometry Using \spitzer\ Priors}

Perez-Gonzalez et al. (2010) have shown that using priors that are measured at shorter wavelengths can 
significantly improve the identification and flux measurement of high-redshift, confusion limited FIR sources. The simplest way to apply
this procedure is to fit two-dimensional gaussians to all detected or suspected sources in the field,  thus increasing the accuracy of the flux
measurement of the prime target. In the case in question,
the main candidates to affect the measurements are lower redshift sources with significant SF that produce some luminosity
due to dust emission even at the long wavelength
SPIRE bands. 

All our nine  detected objects have \spitzer\ observations that  cover the entire SPIRE field of the source. 
We are therefore able to look for neighboring (non-AGN) sources and apply the prior-based flux
measurement method. In eight of the cases the \spitzer\ sources are too far to make a significant effect on the fluxes measured
with the methods described above (\herschel\ and SDSS centered targets). Nevertheless,
we considered the prior-based measurements to be the most reliable and hence they are the ones listed in 
Table~\ref{tab:data}. 
In one case (SDSS J134134) there is an IRAC detected neighbor at 8$''$ from the AGN; well within the SPIRE PSF. Using SDSS-DR9 we identify
this source as a $z=0.53$ galaxy. At such an angular separation, there is no way to use the prior-based method by itself to tell which of the two is contributing more to
the observed FIR flux. We note however that the DR9 spectrum of the galaxy resembles an early type red galaxy with estimated $D_n4000\simeq 1.8$, i.e. very low
specific SFR (sSFR). We also note that assuming most of the \herschel\ flux is from this source, the derived rest-frame SED, in particular flux ratio F(327\mic)/F(163\mic),
 is not consistent with any known starburst template.
 It is more plausible that most or perhaps all of the observed flux is due to the AGN.  

All the six offset sources were observed by \spitzer. Three of those have no nearby neighbors and two others have \spitzer\ detected sources within a distance of 3 \herschel\ PSFs from 
the location of the AGNs. The \herschel\ images
are entirely consistent with these neighbors being the sources of the FIR emission.  Thus all these five sources remain in the offset category which we consider as non-detection.
We did not include these objects in the stacking analysis since there is some residual flux at the AGN location because of the large PSF. 
One of the targets (SDSS J1654) clearly shows a \spitzer\ source near the SDSS location of the AGN. 
The prior analysis of this case results in a satisfactory solution with AGN FIR flux consistent with real
detection. Thus, the final numbers in our sample are 10 detections, 5 offsets and 29 upper limits.

Finally, we note again that while the uncertainties are mostly due
to source confusion, with small additional instrumental noise, the use of priors 
considerably reduce these uncertainties. In the table w keep the listed uncertainties as they are and hence,
in a couple of sources, the measured fluxes  fall below the nominal 
$3 \sigma$ level yet they are  highly significant.

\subsection{\lsf\ measurements}

The assumption we make in determining \lsf\ of individual sources is that almost all the FIR emission is associated
with SF in the host galaxy. This is based on various earlier papers that investigated this issue in detail, e.g \cite{Netzer2007b},
\cite{Schweitzer2006} and \cite{Rosario2012}. The evidence for this is discussed in detail in \S3.2 and 
the following is a description of the procedure used to subtract the residual AGN contribution
to the \herschel\ observation.

To estimate the likely contribution of AGN-heated dust, we assume that this comes from the central torus and made use of the
published \cite{Mor2012a} composite NIR-MIR SED.
The 1--35\mic\ part of this composite spectrum was obtained by fitting a large number
of intermediate luminosity, $z<0.2$ type-I AGNs with a complex model that
includes  a clumpy torus, hot pure graphite dust and dust emission from the narrow line region (NLR). 
The characteristics of this composite  are a sharp rise between 1 and 2\mic, a wide plateau (in terms of $\lambda L_{\lambda}$)
between about 2 and 20 \mic, two broad silicate emission features,  and a significant drop above 20-25\mic. 
We extended the composite to 0.5\mic\ using the data in \cite{Richards2006} and to longer
wavelengths using a 100K gray-body.
\cite{Mullaney2011} carried out a similar analysis using a different fitting procedure that does not
take into account the
NLR dust and the hot dust and employs a different procedure to subtract the SF contribution.
Their mean MIR spectrum is similar to \cite{Mor2012a} up to about 20\mic\ but suggests a drop in $\lambda L_{\lambda}$
at longer wavelengths of about 30--40\mic. 

The \cite{Mor2012a} composite was normalized to the optical-UV continuum of the sources
in their sample.  This composite represents intermediate luminosity type-I AGNs with 
a  total covering factor ($C_f$) of hot (graphite) and warm (torus) dust of $\sim 0.5$. 
For such sources,
$\lambda L_{\lambda} (12 \mu {\rm m} )\simeq 1.25 L_{5100} $, suitable for $f_C=0.5$. 
We also estimated that for our \zzz\ sources, $L_{5100}=0.5 L_{1450}$. 
The combination of all these factors result in  $L_{40 \mu m} \simeq 0.32 L_{1450}$.
As shown in numerous papers \cite[e.g.][and references therein]{Maiolino2007a,Mor2011,Ricci2013}, there is a general tendency for the covering 
factor of the torus to decrease with AGN luminosity, albeit with a large
scatter. This suggests that the assumption of $f_C=0.5$ is likely to overestimate the NIR-MIR emission in our
sample of very luminous objects. Thus, the above scaling of $L_{40 \mu m}$\mic\ relative to $L_{1450}$ is probably an upper limit
to the emission at 40\mic\ and even more so at 43\mic, the rest-frame wavelength corresponding to the 250\mic\ \herschel\ band.

\begin{figure}
\centering
\includegraphics[height=14cm]{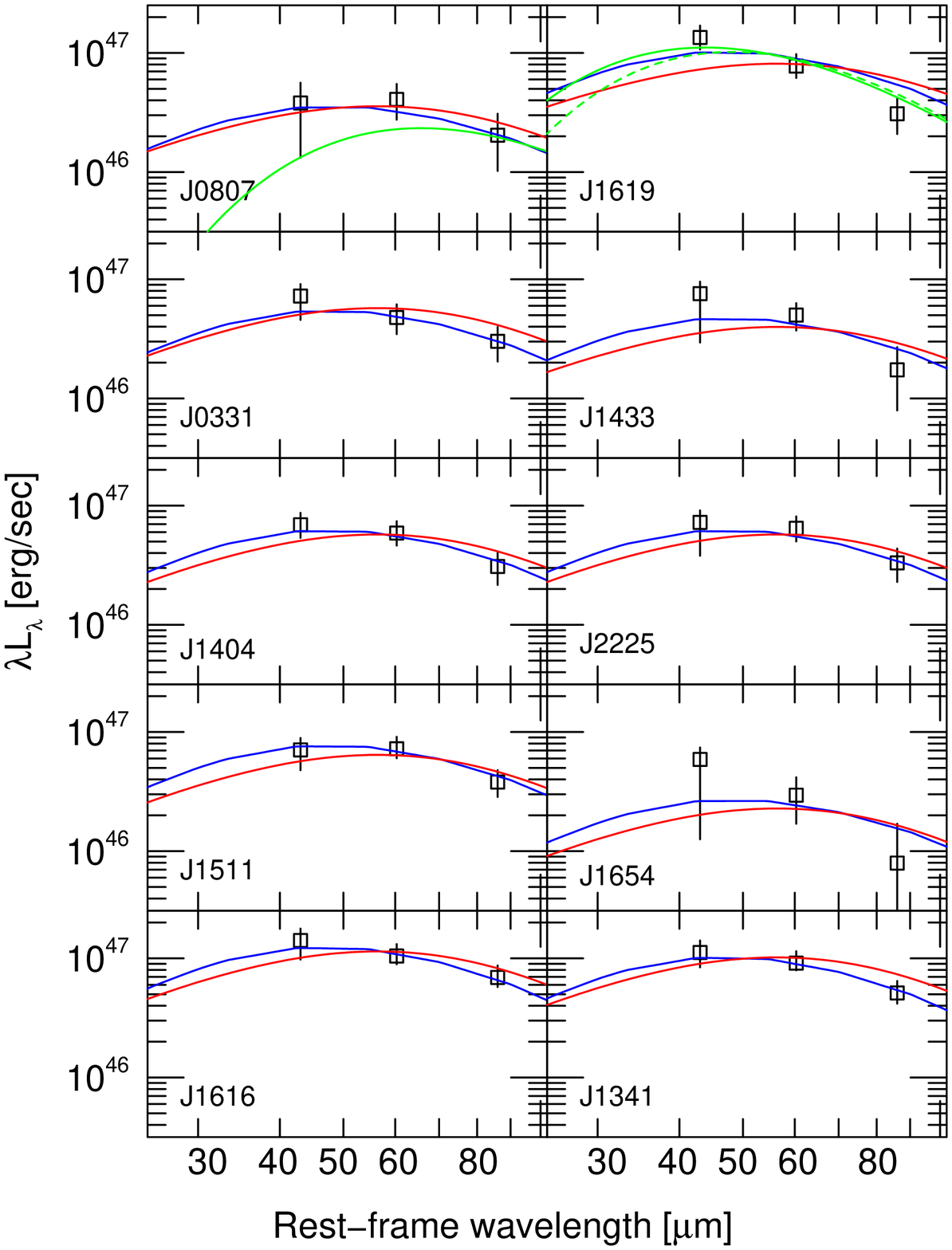}
\caption{
Individual FIR SEDs for the 10 \herschel-detected sources. In all cases the blue curves are templates 
from the \cite{Chary2001} SED library of starburst galaxies and the red curve the composite SED
of \cite{Magnelli2012} for the most luminous SMGs in their sample.
Nine of the 10 sources are fit well with these pure SF galaxy templates.
A T=40K $\beta=1.5$ gray-body fit to the 500\mic/(1+z) point is shown in green in the top left panel 
(see text for more explanation).  Two examples of gray-body fits to the object which could not
be fitted with any of the other templates are shown in the top right panel: 
 $\beta=1.5$ and T=60K (solid green line) and
$\beta=2$ T=50K (dashed green line).
}
\label{fig:individual_seds}
\end{figure}

The 10 individual SEDs are
shown in Fig.~\ref{fig:individual_seds}. For the  uncertainties on individual points, we made
use of the numbers listed in Table 1. 
These are typically small uncertainties, $\sim $20\% and occasionally even smaller, which
reflect only the instrumental noise and the confusion limit.
For the purpose of the fit, we do not allow any of the uncertainties to go below 0.1 dex.
As for the 250\mic\ bin, we added the observational uncertainty from the table and the one due to AGN-heated 
dust in quadrature.
In 3 of the sources (J0807, J1433 and J1654) this makes the uncertainty considerably larger which affects the SED fitting.

The diagram shows several possible fits to the data.
The blue curves are starburst templates from the \cite{Chary2001} SED library of SF galaxies.
Such templates fit very well 9 of the 10 SEDs. The red curve shown in the diagram
is the mean SED for the most luminous SMGs in the 
\cite{Magnelli2012} sample (B. Magnelli, private communication) scaled up or down to fit the data. 
This SED gives satisfactory, albeit not as good fits 
to the same 9 sources We note that the FIR luminosities
of these SMGs are below the luminosities of most of our sources. Given the trend of higher mean dust
temperature with higher \LSF\ discussed in detail below, we suspect that higher luminosity SMG SEDs would be closer
in shape to our sources.  The derived values of \LSF\ using this template are basically indistinguishable from the previous method,
given the uncertainties. We also used the \cite{Magnelli2012} SED to fit only the 350\mic\ and 500\mic\ bands in
order to avoid the shortest wavelength band, where the AGN contribution is more uncertain. These values are typically
0.05 dex below the value obtained by fitting all three bands except for the two cases showing the steep rising
SED (J1619 and J1654) where they are 0.16 dex and 0.1 dex respectively.

The diagram also shows, in green curves, several gray-body SEDs. The top left panel
shows the case of a $\beta=1.5$ T=40K SED fitted to the longest wavelength point ($500/(1+z) \simeq 86$\mic) assuming that at such
a long wavelength the AGN contribution to the heating of the dust is negligible. This SED is in
clear contrast to the \cite{Magnelli2012} SED shown here and 
many luminous starburst SEDs observed by \herschel\ and other missions (see the detailed discussion in \S3.2).
We consider the SFR obtained from such an SED fit to the data, which is typically a factor 2 below the one derived
from our best fits (see column 10 in Table 2), to be well below the real SFR in pure 
starburst systems of similar L(86\mic).
In the top right panel we show two examples of gray-body fits to the only SED which could not be
fitted by the other templates. The cases shown are $\beta=1.5$ and T=60K (solid green line) and
$\beta=2$ T=50K (dashed green line). The two give satisfactory fits to our data given the under-estimated error bars we use.
 As discussed below, the derived mean dust temperatures in the most luminous
starburst systems are also in the same range of temperatures.

As a test of the measured \LSF, we also fitted gray-bodies with various values of dust temperature and $\beta$ the FIR SED
of all the 10 sources. The fit quality is similar to the one obtained from the SF template fit but we prefer not to use it
since there is no real meaning to a single temperature dust.

The formal errors in the \cite{Chary2001} template SED fit is typically very small, of order 25\% or less. However, the
real uncertainty due to the various assumptions on
the SED shape, must be much larger and we estimate them to be 0.2 dex. These more realistic uncertainties are the ones used in the
maximum likelihood analysis described below. 

Finally, we also fitted T=40K gray-bodies to to the 500/$(1+z)$\mic\ band in all 10 sources to get an 
idea about the absolute minimum
SFR which cannot be influenced by the central AGN. Such \lsf\ depend on the assumed value of $\beta$ and for $\beta=1.5$
and are typically a factor 2 below the ones measured
from our other fits. As discussed below in great detail, such dust temperatures are below those measure in pure starburst systems
with SF luminosities that are in the same range as measured here. Raising the dust temperature to 45K, which is more in accord with
high luminosity starburst systems, reduces the mean difference with our fitted values to less than 0.2 dex, similar to the general
uncertainty we adopt in the rest of the analysis. 

Table~\ref{tab:derived_properties} lists all the measured luminosities of all detected sources using the various assumptions. 
We also list $L_{1450}$ from the SDSS spectroscopy,  $L_{3000}$  from our H-band  
spectroscopy, \LAGN\ derived from $L_{3000}$, and \mbh\ and \Ledd\ from T11. 
For the SFR, we use a standard conversion factor based on \cite{Chabrier2003} initial mass function (IMF)
which gives \LSF=$10^{10}$\Lsun\ for SFR of  1\msunyr. 
For this ratio, the SFR of the detected sources ranges
from 1090  to 4239 \msunyr, similar to the values obtained for the most luminous known SMGs \citep[e.g.][]{Riechers2013}

\begin{table}
\begin{center}
\tiny
\caption{Derived Properties}\label{tab:derived_properties}
\begin{tabular}{lcccccccccl}

\hline \hline
%        &      & \multicolumn{4}{l}{\spitzer\ observations$^{\rm b}$} &  \multicolumn{5}{l}{\herschel\ observations$^{\rm c}$} \\
Object & $z^{\rm a}$&$\log\,L_{1450}\,^{\rm a}$&$\log\,L_{3000}$ $^{\rm b}$&$\log\,L_{\rm AGN}$ $^{\rm b}$&$\log\,\mbh$ $^{\rm b}$&
   $\log\,\lledd$ $^{\rm b}$  & $\log\,L_{\rm SF}$ $^{c}$
   & $\log\,L_{\rm SF}$ $^{d}$ &  $\log\,L_{\rm SF}$ $^{e}$ & SFR $^{f}$  \\
(SDSS J) &              & (\ergs) & (\ergs) & (\ergs) & (\Msun) &  & (\ergs) & (\ergs) &(\ergs) & (\Msunyr)      \\
\hline
033119.66-074143.1 & 4.729    & 46.76 & 46.55 & 47.09 & 8.83 &  0.08 & 46.92 & 46.89 (46.82) &46.61& 2174 \\ % 05
080715.11+132805.1 & 4.885    & 46.71 & 46.53 & 47.07 & 9.24 & -0.35 & 46.72 & 46.74 (46.74) &46.44& 1372 \\ % 26
134134.19+014157.7 & 4.689    & 46.87 & 46.73 & 47.26 & 9.82 & -0.74 & 47.17 & 47.14 (47.08) &46.84& 3866 \\ % 28
140404.63+031403.9 & 4.903    & 46.55 & 46.49 & 47.02 & 9.51 & -0.66 & 46.97 & 46.89 (46.85) &46.62& 2439 \\ % 38
143352.21+022713.9 & 4.722    & 47.14 & 46.84 & 47.37 & 9.11 &  0.09 & 46.87 & 46.79 (46.79) &46.37& 1938 \\ % 37
151155.98+040802.9 & 4.670    & 46.62 & 46.32 & 46.86 & 8.42 & -0.26 & 47.02 & 46.94 (46.94) &46.71& 2737 \\ % 41
161622.10+050127.7 & 4.869    & 47.08 & 46.80 & 47.33 & 9.43 & -0.27 & 47.21 & 47.19 (47.14) &46.97& 4239 \\ % 21
161931.58+123844.4 & 4.806$^*$& 46.41 & ---   & 46.91$^*$&---& ---   & 47.18 & 47.19 (47.03) &46.62& 3930 \\ % 20
165436.85+222733.7 & 4.680$^*$& 47.14 & 46.48 & 47.60 & 9.55 & -0.64 & 46.62 & 46.49 (46.39) &46.03& 1090 \\ % 22
222509.19-001406.9 & 4.890    & 46.97 & 46.70 & 47.23 & 9.27 & -0.21 & 46.97 & 46.89 (46.86) &46.65& 2439 \\ % 07
\hline
\end{tabular}
\end{center}
$^{\rm a}$ Derived from the SDSS spectra.\\
$^{\rm b}$ Derived by T11 from $H$-band spectroscopy of the \mgii\ line.
For objects with asterisks, the redshifts were taken from the SDSS DR7 QSO catalog and $L_{\rm AGN}$ was estimated from $L_{1450}$, following a trend based on the samples of \cite{Shemmer2004} and \cite{Netzer2007a}. \\
$^{\rm c}$ Derived by fitting with \cite{Chary2001} models to the \herschel\ fluxes as measured at the SDSS source centers and accounting 
for priors. The uncertainties of these numbers are discussed in the text and assumed to be 0.2 dex in all cases.\\
$^{\rm d}$ Derived by fitting the \cite{Magnelli2012} SMG template to the \herschel\ fluxes, as measured at the SDSS source centers and accounting for priors.
The values in parenthesis are obtained by fitting only the 350\mic\ and 500\mic\ data points (see text).\\
$^{\rm e}$ Derived by fitting a gray-body model with $\beta=1.5$ and $T_{dust}=40$K to $\lambda L_{\lambda}$(500\mic/(1+z)) as described in the text.\\
$^{\rm f}$ Calculated from the \cite{Chary2001} values of $L_{\rm SF}$ assuming a Chabrier IMF (see text). \\
\normalsize
\end{table}

\subsection{Stacking analysis}
\label{sec:stacking}

Twenty nine sources were not detected above 3-$\sigma$ level in any of the \herschel\ bands. Visual inspection of the images shows
no significant emission at the location of the sources and no unusual emission inside the PSF like in the case of the ``offset''
sources. We consider the fluxes measured for these objects as upper limits.
 
To get the average \lsf\ of these sources, we applied a stacking analysis to all their images in each band, as detailed
in M12. We first cut each image to 
a small stamp symmetrically around the center of the pixel that contains the optical location of the source. 
All stamps have an equal number of pixels and are approximately $1' \times1'$ in size. A stacked image is constructed by 
assigning the images with weights according to their respective exposure times, and averaging the images pixel by pixel. 
Since the dominant source of uncertainty is the extragalactic confusion noise, the slightly different exposure times 
have negligible effects on the results. The results of the stacking procedure is a statistically significant signal in all three bands 
that represents the average flux of the individually undetected sources. We refer to this as the ``stacked source'' and measure its flux
in all three bands  in exactly the same way used for the \herschel\ detected sources.

The average values obtained for the stacked source can be biased if the (undetected) sources have a large spread in their intrinsic luminosities. 
In such a case, few sources that are just below the confusion noise limit might skew the result towards a higher average flux. 
To overcome this problem, we used a bootstrap approach to estimate the ``true'' value of the flux of the stacked source and its 
uncertainty.  Out of the 29 ``stamps'' in each band, we choose 10000 random multisets of 29 images where each image may appear more than once. 
We then stacked each multiset and measured the flux of the stacked source. 
 
The averaging of the images assumes that all objects are located at the center of the stamp. Any contribution from neighboring sources 
would be significantly reduced due to the fact that such contributions 
are expected to be randomly distributed in the images. To further minimize the effect of such contributions we randomly 
rotate each image by 90, 180, or 270 degrees before each stacking.  
The final values of the fluxes in each band are taken to be the maximum likelihood values of the corresponding distributions. 
The uncertainties on these values are estimated by measuring the 16th and 84th percentiles, which are assumed to represent the 1-$\sigma$ error. 
Examples of the probability distribution functions (PDFs) of the bootstrap procedure, in all three SPIRE bands, including the definitions
of the maximum likelihood values and the various percentiles, are shown in M12 Fig.~3.   
Finally, the uncertainty on the 250\mic\ luminosity was added in quadrature to the uncertainty due the possible AGN contribution, exactly as
was done for the individual detected sources.
 
The \lsf\ values of the stacked source are calculated by the 
same fitting procedure used to measure the individually detected sources. The best value ($10^{46.23}$\ergs)
corresponds to SFR$\simeq 440$\msunyr\ and is 
obtained from fitting the SED with a \cite{Chary2001} template.
The best $\beta=1.5$ gray-body fit corresponds to T=55K (see Fig.~\ref{fig:median_seds}).
As explained, the contribution due to AGN-heated dust in those sources may be larger which affects mostly the shortest
wavelength band.

To further test the properties of the undetected sources, we divide the 
26 sources with measured \mbh\ into two equal-number bins in each of the following categories: 
\lagn, \mbh, and \lledd. The division is done according to the median value of the property in question. 
For each of the  bins we apply the  stacking analysis described above.
The results are listed in Table~\ref{tab:stacking_1}. There are no significant differences in the mean \lsf\ of the two subgroups in any
of the three variables although the
 two \mbh\ bins show a $\sim 1.5 \sigma$ difference in their stacked 250\mic\ flux.
The difference is such that the lower \mbh\ objects have lower stacked 250\mic\ flux. This may indicate some dependence
of \LSF\ on BH mass among the non-detected objects as found between detected and undetected sources (see \S~\ref{sec:discussion}).
The relatively small number of sources in each of the bins prevents us from further testing this suggestion.

\begin{table}
\begin{center}
\tiny
\caption{Stacking results for undetected \herschel\ sources grouped by their AGN-related properties.}\label{tab:stacking_1}
\begin{tabular}{lcccccccccccccccc}
\hline \hline
   &$F_\nu$(250\mic)&\multicolumn{4}{c}{The 250\mic\ percentiles}   & $F_\nu$(350\mic)&\multicolumn{4}{c}{The 350\mic\ percentiles}     & $F_\nu$(500\mic)&\multicolumn{4}{c}{The 500\mic\ percentiles}     & $\log~L_{\rm SF}$ \\
Group: & (mJy)&0.15&16&84&99.85                 &(mJy)&0.15&16&84&99.85                         &(mJy)&0.15&16&84&99.85                         & (\ergs)  \\
\hline
All non detections  & 5.09 & 2.83 & 4.34 & 6.20 &  7.86 & 5.22 & 2.05 & 4.25 & 6.80 &  9.62 & 4.64 &  0.61 & 3.42 & 5.62 &  7.93 & 46.23\\
\hline
low \LAGN     & 5.83 & 1.26 & 4.21 & 6.63 &  8.67 & 7.41 & 0.94 & 4.84 & 8.74 & 11.85 & 4.00 & -1.98 & 2.05 & 5.48 &  8.23 & 46.30\\
high \LAGN    & 4.59 & 2.29 & 3.85 & 6.28 &  8.66 & 4.15 & 0.57 & 2.86 & 5.61 &  8.83 & 5.01 & -0.65 & 3.44 & 6.74 & 11.21 & 46.23\\
low \mbh      & 4.60 & 0.62 & 3.37 & 5.76 &  7.52 & 5.51 & 1.68 & 3.89 & 7.35 & 11.24 & 5.52 & -3.99 & 2.25 & 7.81 & 12.31 & 46.30\\
high \mbh     & 6.90 & 3.03 & 5.49 & 7.99 & 10.31 & 6.37 & 0.71 & 4.70 & 9.08 & 11.45 & 4.15 & -0.40 & 2.83 & 4.74 &  6.33 & 46.19\\
low \lledd    & 6.22 & 2.95 & 4.97 & 7.25 &  9.80 & 5.88 & 0.21 & 4.27 & 7.92 & 11.10 & 5.12 & -0.13 & 3.36 & 5.76 &  7.65 & 46.30\\
high \lledd   & 5.07 & 0.58 & 3.74 & 6.68 &  8.64 & 6.93 & 1.99 & 4.29 & 8.19 & 11.33 & 3.61 & -4.43 & 1.41 & 6.28 & 11.61 & 46.30\\
\hline
\end{tabular}
\end{center}
\small
The groups are defined following the median values of the sample in the various properties: 
$\log\left(\LAGN/\ergs\right) = 46.81$, $\log\left(\mbh/\Msun\right) = 8.85$ and $\log\left(\lledd\right) = -0.175$
The uncertainties on \LSF\ are dominated by the fitted starburst model and are estimated to be to be 0.2 dex.
\end{table}

\section{Discussion}
\label{sec:discussion}

\subsection{\LAGN\ and \mbh\  groups}

The main finding of the \herschel\ FIR observations is that our sample consists of two distinct
sub-groups, those detected and those that are undetected by \herschel/SPIRE.
In what follows we investigate in detail how these sub-groups are possibly different 
in several other aspects: 
BH properties (\mbh), AGN accretion properties (\LAGN\ and \Ledd) and co-evolution with their host galaxies.

\begin{figure}
\centering
\includegraphics[height=6cm,angle=0]{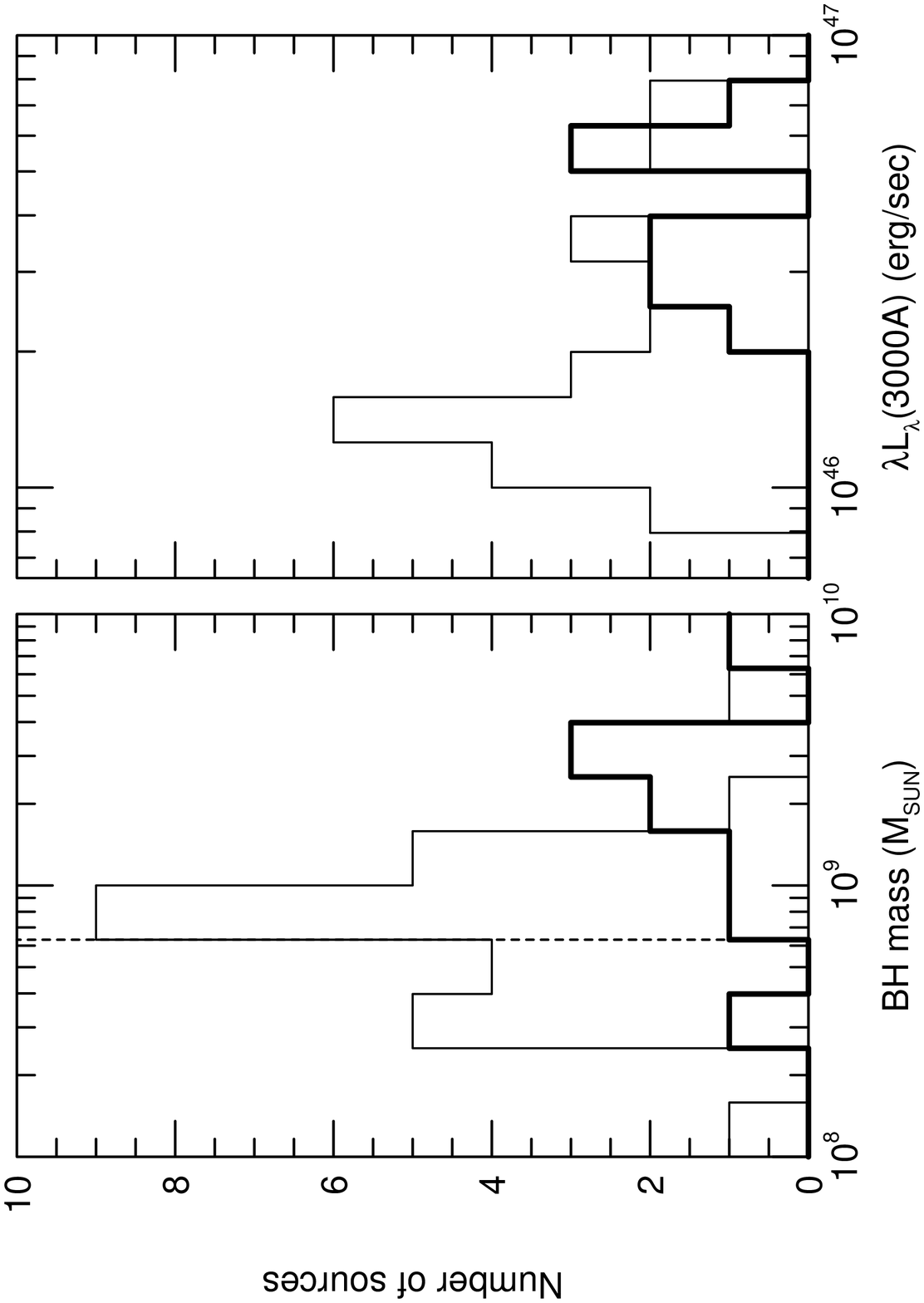}
\caption{
BH mass and $L_{3000}$  distributions showing \herschel-detected (thick line) and undetected (thin line)
sources.
}
\label{fig:mass_histogram}
\end{figure}

In the following we use luminosity proxies like $L_{3000}$ and $L_{1450}$  which are directly obtained
from the H-band and SDSS spectroscopy and do not involve uncertainties related to the bolometric correction factor.
The $L_{3000}$ and \mbh\ distributions of the detected and undetected sources are shown in Fig.~\ref{fig:mass_histogram}.
They exhibit clear separations in both properties; trends that could not be found by M12 because of the smaller sample size 
available at that time (5 detections and 20 non-detections).
We conducted two-sided Kolmogorov-Smirnov (K-S) tests to compare \mbh, various AGN luminosity indicators  and \Ledd\ of 
the detected and undetected sources.
The distributions of \mbh\ for the two sub-groups are significantly different (p=0.005) where the detected \herschel\ sources are showing higher \mbh. 
As for AGN luminosity, the
test shows that the two groups are significantly different (p=0.006) with the detected sources having the higher $L_{3000}$.
A similar analysis using $L_{1450}$ instead of $L_{3000}$ gives very similar results.
Finally, the K-S test for  \Ledd\  shows no significant difference between the groups.

\begin{figure}
\centering
\includegraphics[height=6cm,angle=0]{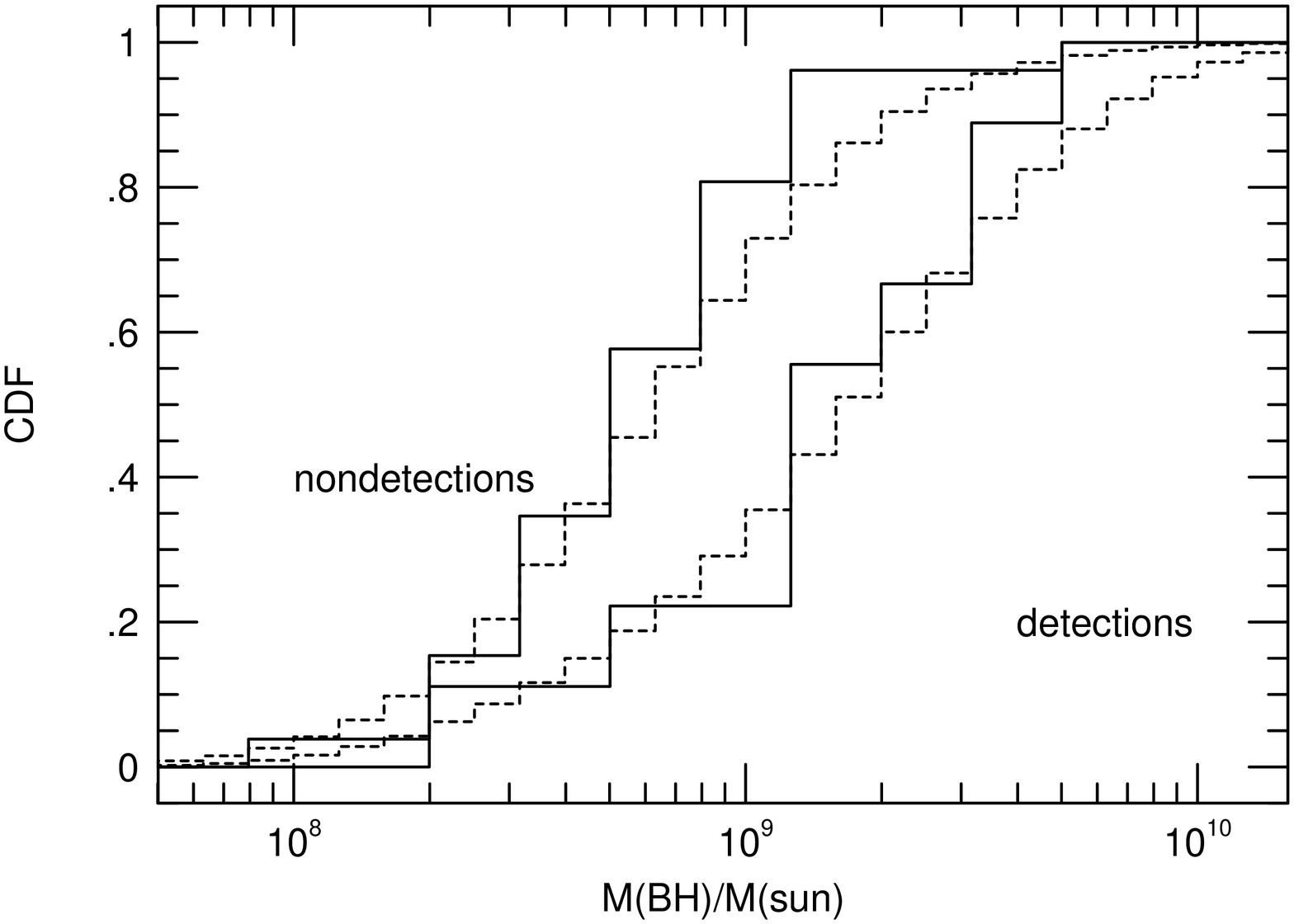}
\caption{CDFs of  \herschel-detected and undetected sources. The dashed lines show the results of the resampling MC simulations.
}
\label{fig:BHM_CDF}
\end{figure}

Next we consider the mean properties of the two groups. Since the groups are small, and the uncertainties quite large,
we prefer to use a statistical method that takes into
account the individual uncertainties on mass and luminosity and the cumulative distribution functions (CDFs) of these properties. 
We assigned to all objects a 0.3 dex uncertainty on \mbh\ and \Ledd\ 
and a 0.1 dex uncertainty on $L_{3000}$ and $L_{1450}$.
Regarding \LSF, similar to the case of individually detected sources, the uncertainties are mostly due to the choice of the SF template
and are estimated to be 0.2 dex.
We use the 
16th and 84th percentiles from the stacking analysis of the 350\mic\ images and added to it, in quadrature, the 0.2 dex due to the model uncertainty.
We then used a bootstrap resampling Monte-Carlo (MC) method, with 10000 simulations, that allows us to draw 
every object many times assuming all uncertainties are normally distributed.
The results are the 16th, 50th and 84th percentiles of the distributions
(note again that we prefer this over a simple mean because we are after the range of properties represented in these
groups rather than the scatter in properties).
For high accuracy measurements, the 50th percentile is almost identical to the median. At large uncertainties, however, the two can differ 
substantially and the resampled  sample follows the CDF much better.
 The values obtained in this way are listed in Table~\ref{tab:most_likely} and the two  CDFs for the BH mass in the groups  
are shown in Fig.~\ref{fig:BHM_CDF}

Before continuing the analysis we comment on the \mbh-\LAGN\ correlation and its dependence on the method used
to obtain BH masses. We are using the ``virial'' (single-epoch) mass estimate method which, for the sample
in questions, assumes \mbh$\propto L_{3000}^{0.62}$FWHM(\MgII)$^2$. The justification and rational for this is approach are detailed
in numerous earlier publications, most recently in \cite{Trak2012}. We checked the distribution of FWHM(\MgII) against
$L_{3000}$ in our sample and found no significant correlation between
the two. Thus, it can be argued that the separation into two \mbh\ groups is a direct result of the assumed dependence of
\mbh\ on $L_{3000}$. This suggestion is not in conflict with the basic premise that the emissivity- weighted size
of the broad emission line region (BLR) is controlled by the radiation field of the source and we are looking at a virialized system. Under such conditions,
there is indeed a direct
connection between the central BH mass and \LAGN. 
Thus, for a sample where FWHM(\MgII) is independent of $L_{3000}$, \mbh\ indeed scales with the source luminosity.
 
We carried out a similar maximum likelihood analysis to obtain the mean observed fluxes
of the groups in the two \spitzer\ bands, two \wise\ bands 
and the three \herschel\
bands. These values are the ones used in the SED analysis discussed below.

\begin{table}
\begin{center}
%\tiny
\caption{Median values and the 16th and 84th percentiles (marked as lower and upper errors, respectively)
of various physical properties.
}
\begin{tabular}{lcc}
\hline
\hline
Property                          & 9 detected sources       & 26 undetected sources\\
\hline
$\log \LSF$ (\ergs)               & $46.93^{+0.25}_{-0.25}$   & $46.23^{+0.31}_{-0.22}$   \\
$\log L_{1450}$ (\ergs)           & $46.84^{+0.28}_{-0.24}$   & $46.49^{+0.23}_{-0.28}$ \\
$\log L_{3000}$ (\ergs)           & $46.60^{+0.21}_{-0.20}$   & $46.25^{+0.24}_{-0.21}$\\
$\log L_{AGN}$ (\ergs)            & $47.13^{+0.26}_{-0.25}$   & $46.78^{+0.26}_{-0.25}$\\
$\log$(\mbh)  (\msun)             & $9.28^{+0.45}_{-0.52}$    & $8.85^{+0.39}_{-0.38}$ \\ 
$\log$(\Ledd)                     & $-0.329^{+0.51}_{-0.48}$  & $-0.18^{+0.44}_{-0.40}$ \\
\label{tab:most_likely}
\end{tabular}
\end{center}
$^{\rm a}$ 
The assumed uncertainties on individual measurements are: for detected sources,  
0.3 dex on \mbh\ and \Ledd, 0.1 dex on $L_{3000}$ and $L_{1450}$, and 0.2 dex on 
\LAGN\ and \LSF. For \LSF, this is larger than in Tables 2 \& 3 since it reflects also the uncertainty on the model rather
than the fitting procedure itself. For undetected sources,
same uncertainties on all quantities except for \LSF\ where we used the 
 percentiles obtained for the 350\mic\ stacked image together with the individual 0.2 dex uncertainty. 

\end{table}

\subsection{Broad band SEDs}
\label{sec:SED}

\subsubsection{Median spectra and global SED shape}

We compared the mean monochromatic luminosities of the two groups across the entire wavelength range available to us: 
SDSS spectroscopy corresponding to rest-frame 900--1630\AA,  
H-band (rest-frame 2530--3160\AA) and K-band (rest-frame 3480--4310\AA) spectroscopy from 
T11, \spitzer\ photometry (effective rest frame wavelengths of 6120 and 7760\AA),
\wise\ photometry (effective rest-frame wavelengths of 5780\AA, 7758\AA, and 1.99\mic) 
and the \herschel\ photometry. The objects used in this analysis are only those with measured BH mass and \spitzer\ observations,
i.e. 9 detected and 24 undetected sources. We combined all the available information for all objects in both groups  
 even if they do not have K-band data\footnote{Of the 40 objects in T11, 24 have K-band spectroscopy. Six of those are 
\herschel-detected with \spitzer\
observations, 17 \herschel-undetected with \spitzer\ observations, and one \herschel-undetected with no \spitzer\ data.}.

We experimented with three different ways of combining
the optical data: median spectra, simple mean and geometrical mean (averaging their $\log \lambda L_{\lambda}$). 
We prefer the geometrical mean that is not biased by
a small number of outliers (in terms of luminosity) and, at the same time, does a better job in reducing the noise of
individual measurements. The geometrical mean and median spectra are quite similar but the signal-to-noise (S/N) in the first
is somewhat higher.
For the \spitzer\ bands, and the 2-\wise\ short wavelength bands, we used the 
resampling MC method to estimate the 50th percentile fluxes in both groups.
 For the longer wavelength
\wise\ band (11.6\mic) we used the same method combined detections and non-detections.
This results in a significant 11.6\mic\ detection for the group of 9 \herschel\ detected sources and a point which is consistent with L(11.6\mic)=0 in the other group.

As for the \herschel\ observations, we tried a variety of model SEDs.
For the detected sources excluding 161931.58+123844.4 (which does not have a BH mass measurement),
we used the maximum likelihood method with its 16th and 84th percentiles and included
the possible AGN contribution to the 250\mic\ band. This results in a statistically significant fit with \LSF=$10^{46.93}$\ergs.
The \cite{Magnelli2012} SED with the same \LSF\ also goes through the three points. 
We can also obtain satisfactory fits for a T=60K gray-body with \LSF=$10^{47.0}$\ergs\ or lower temperature larger $\beta$ gray-bodies
with very similar \LSF..
Finally, a $\beta=1.5$ T=40K gray-body fit to the 500\mic\ band, as attempted for the individual sources,
 gives an unsatisfactory fit with \LSF=$10^{46.63}$\ergs.

For the \herschel-undetected sources, 
we used the results of the stacking analysis but increased the uncertainty on the 250\mic\ flux according to the assumed AGN contribution.
This resulted in a satisfactory SF template fit with \LSF=$10^{46.19}$\ergs. This value is $\sim10$\% lower than the one obtained
for the case with no AGN contribution lower than the one obtained
for the case with no AGN contribution. A $\beta=1.5$ T=55K gray-body with \LSF=$10^{46.15}$\ergs\ also provides a good fit. Finally,
the T=40K gray-body fit to the 500\mic\ flux gives an unsatisfactory fit with \LSF=$10^{45.95}$\ergs.

\begin{figure}
\centering
\includegraphics[height=8cm]{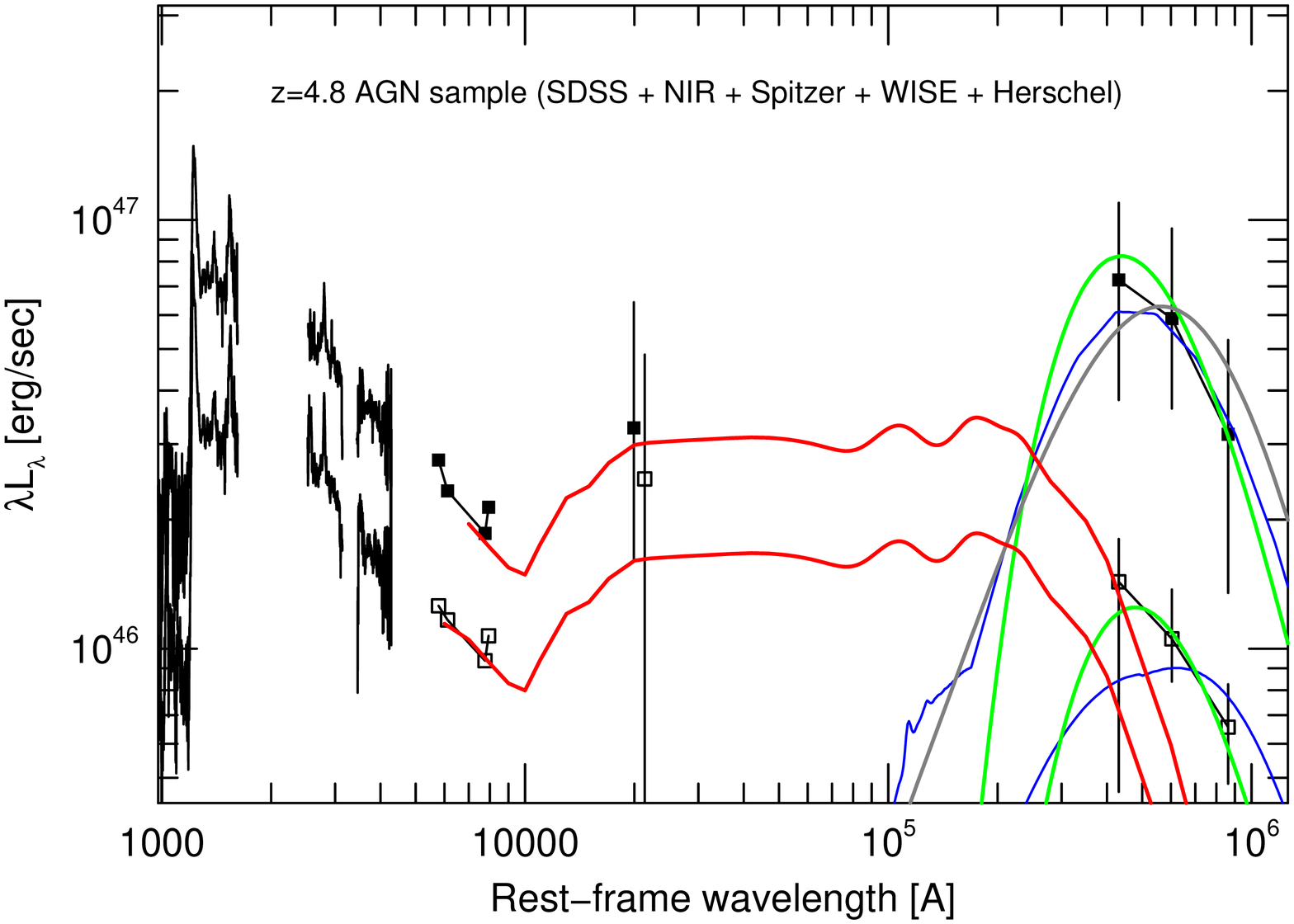}
\caption{
Mean (maximum likelihood) SEDs for \herschel-detected (upper spectrum, 161931.58+123844.4 not included since it has no BH mass
measurement) and undetected (lower spectrum) sources. The red curve is the \cite{Mor2012a} 
NIR-MIR composite normalized to the 50th percentile value of $L_{3000}$ and extended to shorter and longer wavelengths (see \S2.6). 
The diagram shows  two FIR fits for each of the \herschel\ groups, template SF SED in blue and $\beta=1.5$ gray-body in green:
For the detected sources it also shows, in gray, the \cite{Magnelli2012} template normalized to the same \LSF\ as the template SF SED.
For the detected sources, \LSF=$10^{46.93}$\ergs\ for the template SF SED and \LSF=$10^{47.0}$\ergs\ for
a T=60K gray-body.
For the stacked undetected sources, the template SF SED gives \LSF=$10^{46.19}$\ergs\ (i.e. smaller by 10\% from our fitted stack spectrum that
does not include the possible AGN contribution) and the gray-body has T=55K and \LSF=$10^{46.15}$\ergs.
The rest-frame 1.99\mic\
\wise\ points are shifted slightly in wavelength, for clarity.
}
\label{fig:median_seds}
\end{figure}

The various SEDs discussed above are shown in Fig.~\ref{fig:median_seds}. The SF template SEDs are plotted in blue
and the gray-body SEDs in green. For the detected sources we also show, in gray color, the \cite{Magnelli2012} SED normalised
to the same \LSF\ as the template SF SED. 
We also plot the  normalized mean MIR spectrum from
\cite{Mor2012a} extended to shorter and longer wavelengths as described in \S2.6.

The fitting of the FIR data of individual \herschel-detected sources was discussed in \S2.6 and illustrated in Fig.~\ref{fig:individual_seds}.
Here we comment on the general trend of increasing mean dust temperature with \LSF\ and the distance from the MS 
as reflected in our sample. The discussion is more relevant to the overall shape of the SED, and its peak, since fitting 
the FIR observations by a single temperature gray-body is obviously over simplified.

Recent studies of \spitzer-detected and \herschel-detected high SFR sources established the great similarity of the FIR SED of 
SF galaxies, starburst galaxies and AGNs of the same FIR luminosity. They also show the general trend of increasing mean dust temperature
with increasing SFR. 
Schweitzer et al. (2006) and various papers since used \spitzer\ data to show that the FIR SEDs and L(PAH)/L(FIR) are both very similar
in AGNs (mostly PG quasars) and ULIRGs of the same redshift and FIR luminosity. This suggests the SF origin of the FIR emission in these
AGN-dominated and SF dominated sources.
Two more recent studies, by \cite{Magnelli2013} and \cite{Symeonidis2013}, focus on typical SF galaxies (mass sequence galaxies, MS, see
more details in \S~\ref{sec:mass_sequence})
at all redshifts up to $\sim 2$. They 
demonstrate the steady increase in mean dust temperature with SFR reaching T$>40$K at high SFR and high redshift (that are still well below the
redshift and luminosity in our sample).
Regarding objects above the MS, thought to be powered by powerful starbursts whose origin is related to major
galaxy mergers, the increase in mean dust temperature is even more noticeable. For example, \cite{Magnelli2012} studied a large
group of SMGs showing that the mean dust temperature is increasing with \LSF\ reaching peak luminosity ($\lambda L_{\lambda}$)
at around rest-frame wavelength of 55\mic. This can be translated to dust temperature larger than 50K (depending on the assumed $\beta$)
at the highest luminosity end
where \LSF$\simeq 10^{46.8}$\ergs, about 0.15 dex below the median FIR luminosity of our \herschel-detected sources.
Study of SMGs at somewhat higher redshifts and similar luminosities by \cite{Roseboom2012} reaches similar conclusions.
Finally we note that \cite{Magdis2010} also found very high dust temperature, reaching 60K, in a sample of $z\sim2$ ULIRGs but in
this case the trend with FIR luminosity is much weaker.

Out of the 11 SMGs in the \cite{Magnelli2012} group of the highest FIR luminosity sources, two show weak AGN 
emission  with \LAGN\ two orders of 
magnitude below the AGN luminosity in our sample. There is no way such AGNs can contribute more than a few percent to the FIR
luminosity of these objects.
The 9 other SMGs show no hint for an AGN emission confirming, again, the suggestion of SF-heated dust. As 
Fig.~\ref{fig:individual_seds} shows, a scaled version of the
Magnelli et al. SED, fits  well the FIR SEDs of our \herschel-detected sources allowing for the hot AGN dust contribution to
the shortest wavelength point in three of the sources (see details in \S2.6). 
A 50K gray- body with $\beta=2$ and a 60K gray-body with $\beta=1.5$ also provide adequate fits to the mean
spectra shown in Fig.~\ref{fig:median_seds}.

There is further confirmation of high temperature dust, with FIR SED similar to the one observed in our sample, in several
high SFR, high-z sources that do not show any sign of AGN activity. Examples are a $z=6.34$ sources found by
\cite{Riechers2013} (T$\sim 56$K) and several lensed starburst systems studied by \cite{Weiss2013}, with
$z\sim 4.5$ and T$>50$K.
Thus the FIR SEDs of our \herschel-detected sources allowing for the small hot AGN dust to the 250\mic\ bin in 3-4 of the sources,
 and the derived peaks of the  rest-frame $\lambda L_{\lambda}$,
are indistinguishable from
those of similar FIR luminosity sources with no sign of AGN activity. 

Finally, as explained in the previous section, this may not be the case in all of the undetected sources. In some
of those, most likely the fainter ones, AGN-heated dust can contribute more to the 250\mic\ and perhaps even the  350\mic\ 
emission. This possibility cannot
be ruled out given only a single stacked spectrum. The limit obtained in the previous sections 
from fitting the stacked 500\mic\ flux 
with a 40K gray body (SFR$\sim 230$\msunyr, see explanation to Fig.~4) suggests that the uncertainty on the SFR of the stack image due
to such contribution is a factor of 2 at most.

\subsubsection{Accretion disk models}
\begin{figure}
\centering
\includegraphics[height=8cm]{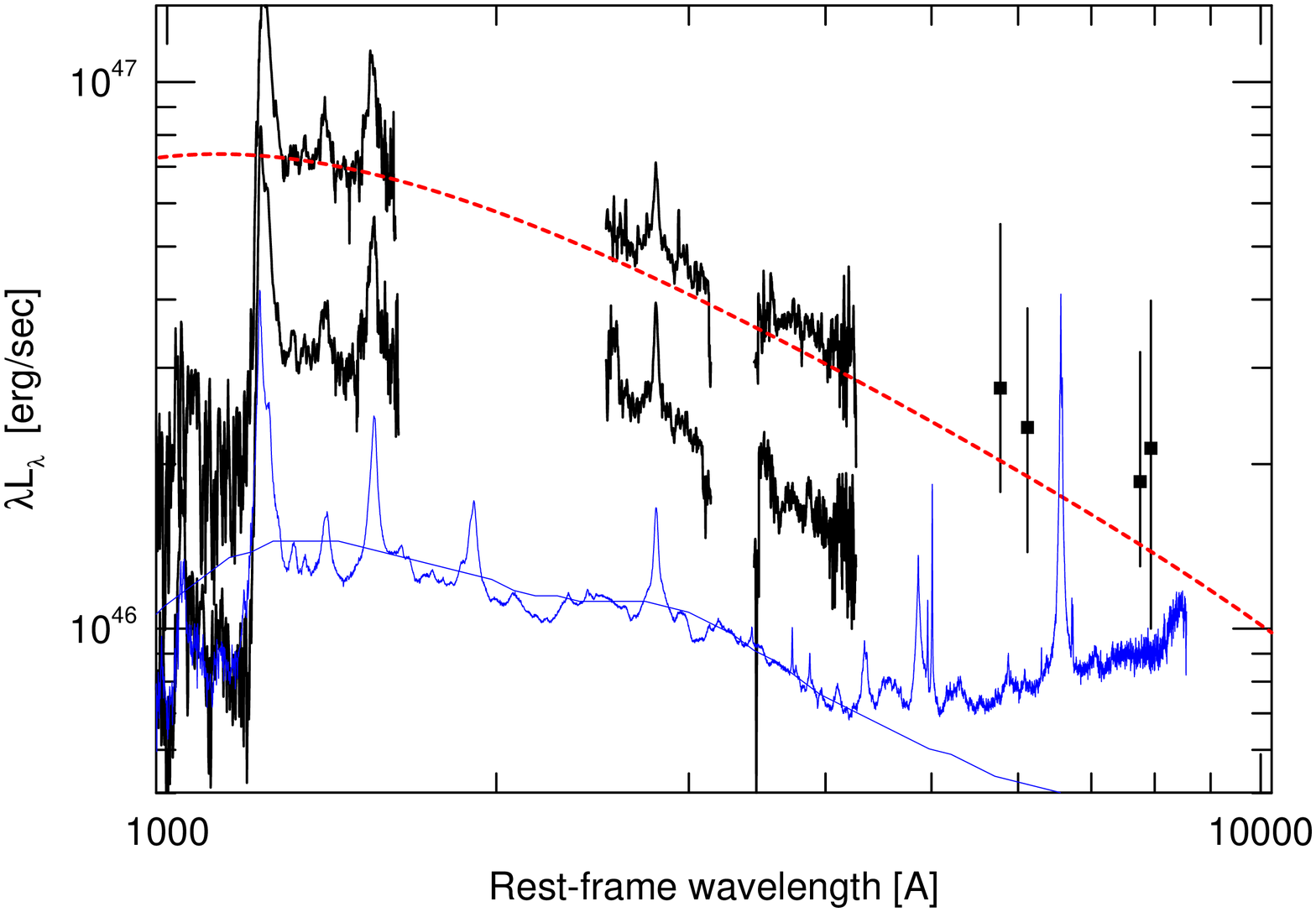}
\caption{
Expanded view of the SEDs in Fig.~\ref{fig:median_seds} (upper spectrum - detected sources, lower spectrum - undetected
sources without the \spitzer\ and \wise\ data points)
showing the comparison with two AGN composites and an accretion disk model.
The composites are from 
\cite{vandenberk2001}  (high resolution blue curve) and \cite{Richards2006} (smooth low resolution blue curve).
The \spitzer\ 3.6\mic\ flux was reduced by 25\% to account for the strong \ha\ line  included in this band.
The mean observed SED of the \herschel-detected sources (top black spectrum) is very similar to what is expected from a thin accretion disk which
is shown by the dashed red line. The disk model assumes \mbh$=2 \times 10^9$\msun,
spin parameter of 0.2 and $L/L_{Edd}=0.5$, very similar to the mean measured properties of this group.
}
\label{fig:expanded_sed}
\end{figure}

To further investigate the short wavelength parts of the SEDs, we compared them to known AGN composites and to several theoretical
accretion disk models. To make a more meaningful comparison, we subtracted the estimated contribution of the \ha\ line from the
flux in the \spitzer\ 3.6\mic\ band by assuming that the line contributes 25\% of the flux in this band \citep{Stern2012}. 
An expanded view of the modified SEDs is shown in Fig.~\ref{fig:expanded_sed}.
The composite AGN spectra shown in the diagram (blue lines) are taken from 
\cite{vandenberk2001} and \cite{Richards2006}. The first one is known
to represent well the 1000--5000\AA\ spectrum of many intermediate luminosity AGNs. It drops very steeply shortward of the \La\ line because of
the inclusion of many high redshift sources that are affected by inter-galactic absorption,
 and flattens longwords of about 5500\AA\ because of the host galaxy contribution in the spectrum of lower luminosity,
lower redshift sources. The second is representative of more luminous SDSS AGNs hence the galaxy contribution is less noticeable.
As seen in the diagram, both composites are
considerably different from the mean SEDs of the \zzz\ sources studied here over the common rest-frame  wavelength.

A plausible origin of the UV-optical continuum of luminous AGNs is emission by optically thick, geometrically thin accretion disk. This has been studied in numerous
papers (see \cite{Blaes2007} for a review) yet the comparison with the observations is 
still ambiguous. In particular, there are many low and intermediate luminosity AGNs where the observed, galaxy subtracted SED deviates considerably from
the predicted thin accretion disk spectrum. There are also large uncertainties to do with the host galaxy contribution at long wavelengths.

The \zzz\ sources studied here are very luminous and reside in host galaxies with high SFR and
very little, if any, old stars that contribute significantly to the spectrum  
around 0.5--1\mic. This  assures a dominant AGN contribution over the entire rest-frame range
of 1000--8000\AA\ and enables
a much cleaner comparison with accretion disk models.
 To check this more quantitatively, we calculated several disk spectra using the code
developed by \cite{slone2012} who studied disk SEDs in the presence of fast disk outflows
(the results shown here do not include disk winds).
The face-on spectrum of one such disk model, with no mass
outflow,  \mbh=$2 \times 10^9$\msun, $a=0.2$ ($a$ is the BH spin) and \Ledd=0.5, is shown in Fig.~\ref{fig:expanded_sed}.
The theoretical spectrum is in very 
good agreement with the mean SED of the \herschel-detected sources whose mean \mbh\ and 
\Ledd\ are in turn very similar to the model parameters, except for the unknown spin.
To the best of our knowledge, no previous AGN SED study shows such a good agreement between observed and computed accretion disk spectra. For example, the recent work
of \cite{Landt11} makes a detailed comparison of this type for several low luminosity AGNs but their fits are  model dependent because of the various assumptions
about the host galaxy contribution and the unknown rest-frame UV part of the spectrum.
Additional information on the long wavelength spectrum of AGN disks can be found in \cite{kishimoto2007} who studied the polarized spectrum 
of several objects. The long wavelength parts of these spectra are similar to the predicted $\nu^{1/3}$ disk spectrum  but, again,  these data do not extend to short enough  wavelength
to verify the predicted flattening part of the disk SED. The very high redshift sources in our sample, and the combined SDSS and H and K-band spectra, provide
a unique opportunity to study such disks over a very wide wavelength bands and suggest a very good agreement with the models.

The above comparison is not meant to address in detail the optical-UV spectra  of high BH mass AGNs (which is the subject of a forthcoming paper).
The purpose is to check two {\it mean SEDs} with a considerable range of properties within each group. 
One should also be aware of the fact that the high \Ledd\ of most of the sources
may be in odds with the conditions normally assumed for thin accretion disks.

\subsection{Stellar mass BH mass and star formation at \zzz}
\label{sec:mstar_mbh}

\subsubsection{\LSF\ \LAGN\ and \mbh\ at \zzz}

The main premise of our work is that the \zzz\ sources represent the parent population of the most
massive BHs, those reaching a mass of $few \times 10^{10}$ \msun\ in the local universe. Such objects have
now been observed in a small number of local galaxies associated with rich and massive galaxy clusters \cite[e.g.][and references therein]{mcconnell2012}.
As demonstrated in T11 (Figs. 6 and 8 there), this assumption connects the \zzz\ population to the more massive
BHs that power the most luminous AGNs 
at $z\simeq 3.3$ and $z\simeq 2.4$ discussed in detail in T11. 
Under this hypothesis, additional BH growth from \zzz\ to $z\simeq 3.3$ (about
$7 \times 10^8$ yr) and  $z\simeq 2.4$ (about $1.5 \times 10^9$ yr) requires a mean duty cycle of 0.1--0.2 if the growth is linear
(constant \LAGN) and about 0.05--0.1 if it is exponential (constant \Ledd).

A major finding of the present work is the link between \LSF, \LAGN\ and \mbh. 
This was not found by M12 who analyzed a smaller sub-sample that included 5 detected and 20 undetected sources. 
M12 considered various possible
scenarios to explain the properties of the \zzz\ sources.
Their preferred explanation included a population of sources with similar BH mass and AGN luminosity. In this scenario, the undetected sources are those that already
finished their most active phase, related to a merger of large, similar mass galaxies, and are starting their post-starburst phase, possibly in secularly evolving systems. 
According to this explanation, the mean \mbh\ in the group of undetected sources should equal or exceed
the mean \mbh\ of the detected sources, in clear contrast with the present finding.

Here we propose two alternative scenarios that are different from the ones presented by M12, and are related to the presence or absence of AGN feedback.
These are by no means the only possibilities and the reader is referred to other papers, \citep[e.g.][]{Aird2012,Bongiorno2012}, addressing the
connections between BH and stellar masses, albeit at much lower redshifts and much less robust determination of \mbh\ and \Ledd.

Both explanations suggested here are based on the assumption that all the objects in our sample reach \zzz\ with basically the same 
properties like stellar mass, BH mass and accretion rate.
The first explanation assumes that 
AGN feedback is very rare or inefficient in massive objects, like those in our
sample, at \zzz, and does not affect the AGN and stellar
mass evolution. The suggestion is that the active BHs reside in galaxies
of different SFRs, possibly because of various ongoing interactions. For BHs that are hosted in lower SFR galaxies,
the amount of cold gas available for SF and for BH accretion is smaller than in the other, higher SFR objects.
This results in lower \LAGN\ and slower BH growth.
Other objects are hosted in systems where cold gas is more abundant. Such objects increase both their stellar and BH mass at a faster rate. 
Over time, the mass of the BHs in 
the second group will considerably exceed the mass of the BHs in the first group. Assuming uninterrupted, BH exponential growth, and using the 
median values of \mbh\ listed in
Table~\ref{tab:most_likely}, we can estimate that the time requires to establish the mass difference between the two groups is $\sim 6 \times 10^7$ yr. 
This is shorter than the difference in cosmic time between the lowest
and the highest redshift objects in our \zzz\ sample.

The second possibility requires AGN feedback in some but not all objects. The differences depend on the local conditions, most
likely the geometry of the interaction. 
In this scenario, all objects start in galaxies that undergo violent mergers, and extreme SF, with rates comparable to the ones measured in the 
group of \herschel-detected sources.
The geometry of the merger, in particular the location of the cold gas is such 
that in most objects the AGN activity affects the surrounding gas, 
either by photoionization
or by fast outflowing winds (mechanical feedback). The result is a large decrease in SFR,  by a factor of about 5 (the difference in the mean SFR of the two groups). 
The majority of the objects in our sample (75\%) belong to this group. 
The remaining 25\%\  are different,
possibly due to the different geometry, and/or the mass outflow rate in the vicinity of the BH.
These sources undergo no feedback.
One can also consider variants of these scenarios, e.g. that all sources start in systems 
with steady SFR but
some of them undergo a major merger at later times which raises both their SFR and BH accretion rate. 

Both scenarios considered here link BH and stellar mass growth to the cold gas supply to the system. This naturally explains the connections between \LSF, \LAGN\ and \mbh.
We note that the feedback assumed in the second scenario is not efficient enough to shut down completely
all the SF or BH accretion in those cases where it has an effect on the evolution of the system. 
For example, it can shut down SF in the central region but does not affect SF in the
outer extended disk.
Obviously, we cannot exclude the possibility that similar mass BHs at \zzz\ undergo very efficient feedback that results in dormant BHs in quenched galaxies that cannot be observed at that redshift.
Such objects are not present in our sample.

\subsubsection{The mass sequence and \mstar/\mbh\ at \zzz}
\label{sec:mass_sequence}

The terminology used here is drawn from  studies of  lower redshift SF galaxies, in particular
their distribution in the SFR vs. \mstar\ plane.
We focus on the mass sequence (or main sequence, MS) introduced first by \cite{Noeske2007}, \cite{Daddi2007}, and several others. 
We assume that such a sequence is present also at very high redshift.
As shown in the earlier works, and in numerous recent papers \cite[e.g.][]{Rodighiero2011,Santini2012,Wuyts2011a,Wuyts2011b,
Rosario2012,Whitaker2012}, for a given \mstar,  the MS SFR increases with redshift up to about
$z \sim 2.5$ where it flattens and possibly even drops with increasing redshift. The evolution beyond $z \sim 2.5$ is still
a matter of debate with some claims that observational issues such as contamination by strong emission lines, might have affected
the earlier results \citep{Stark2013}.

The highest observed \mstar\ at high redshift  differs slightly
from one study to another depending on the measurement method and various systematic effects.
For example, \cite{Wuyts2011a} and \cite{Wuyts2011b} show that at 
$z \sim 3$, \mstar=$10^{11-11.5}$\msun\ which corresponds to SFR$\simeq 300-1000$\msunyr.
Other papers give, for a similar redshift and \mstar=$10^{11.5}$\msun,
lower SFR of about 400 \msunyr.
 The largest stellar masses at $z \sim 4$, where such measurements  are still feasible,
are about $10^{11.6}$\msun\ \cite[e.g.][]{Marchesini2009}).  
There are several SFR and \mstar\ estimates at even higher redshift, 
\cite[e.g.][ and references therein]{Lidman2012}, 
but these are for much lower stellar mass galaxies.
Numerical simulations, like the ones reported in \cite{Khandai2012},
 are inconclusive about this issue with some suggestion
that the largest stellar mass beyond $z=4$ cannot exceed $\sim 10^{11}$\msun. In the following, we follow the assumption that the MS at \zzz\ is similar
to the one observed at $z \sim 2.5$ but note that the numbers derived 
under this assumption for \mstar\ may well represent upper limits.
The SFR for the assumed MS is about 100\msunyr\ for \mstar=$10^{10.5}$\msun\ and the slope in the range 0.5-0.7..

There are several possibilities regarding the location of the \zzz\ sources relative to the MS (which we assume exists at this redshift),
but none can be proven at this stage.
One possibility is that {\it all} sources are on the MS with unknown stellar masses or with stellar mass at the high mass end mentioned above.
Regarding the \herschel-detected sources, this idea seems to be in conflict with the suggestion of little or no evolution at $z>2.5$ since at
lower redshift, galaxies with SFRs as high as measured by us are located well above the MS. Moreover, as shown in 
\cite{Riechers2013}, non-active galaxies at $z>4$ with SFR similar to those of the detected sources,
show clear signs of starburst activity. 

Another possibility is that the undetected sources
are on the MS and the detected ones are in starburst, possibly merging systems above the MS, as is often seen at lower redshift
\citep[e.g.][]{Rodighiero2011,Sargent2012}. Given the problematics of the first hypothesis, we only explore the consequences of the latter 
case combined with the assumption that \mstar\ is 
as discussed above, i.e. the highest possible at this redshift. The unknown stellar mass at \zzz, does not justify, 
in our opinion, to explore a  larger range of possibilities. 

Our best estimate of the mean SFR in the group of 29 \herschel-undetected sources is 400--440\msunyr.
Assuming these are all MS source with \mstar$\sim 10^{11.5}$\msun.
and using the median measured value of \mbh\ for this group
 ($10^{8.85}$\msun), the mean \mstar/\mbh\ is about 450. The uncertainty on this number is at least
as large as the one derived from the mean width of the MS at high redshifts
 \cite[e.g.]{Whitaker2012,Rodighiero2011}, which is about $\pm 0.3$~dex.
 As for the detected sources, if their stellar mass is the largest at this redshift, i.e. similar to the stellar mass of the 
undetected sources,
 we can thus guess that their mean
\mstar/\mbh\ is smaller by about a factor 3, similar to the difference in the mean
\mbh\ of the two groups. This translates to \mstar/\mbh$\sim 150$. The entire range of \mstar/\mbh, 150--450, is similar to the ratio observed
in the local universe for the most massive BHs.

\subsection{Black hole and stellar mass evolution from \zzz\ to \z24}

 The larger sample size, and the division into two \LSF, \mbh\
and/or \LAGN\ groups, allow us to consider in some detail the instantaneous and cumulative growth of the BH and stellar masses in our sample.

\subsubsection{Instantaneous BH and stellar mass growth: \LSF\ vs. \LAGN}

As already mentioned in \S1, the correlation (or lack of) between \LSF\ and \LAGN\ has been debated extensively in the literature, see e.g. 
\cite{Netzer2007b,Netzer2009a,Shao2010,Rosario2012,Harrison2012,Page2012}.
The work of \cite{Rosario2012} summarizes many of the earlier results and used \herschel-based \LSF\ measurements to illustrate 
the redshift dependence of this relationship
and the transition from no correlation to a fix slope correlation typical of AGN dominated sources (\LAGN$>$\LSF) described in \cite{Netzer2009a}.
Several attempts to explain this behavior are discussed in \cite{Neistein2013}.

\begin{figure}
\centering
\includegraphics[height=6cm]{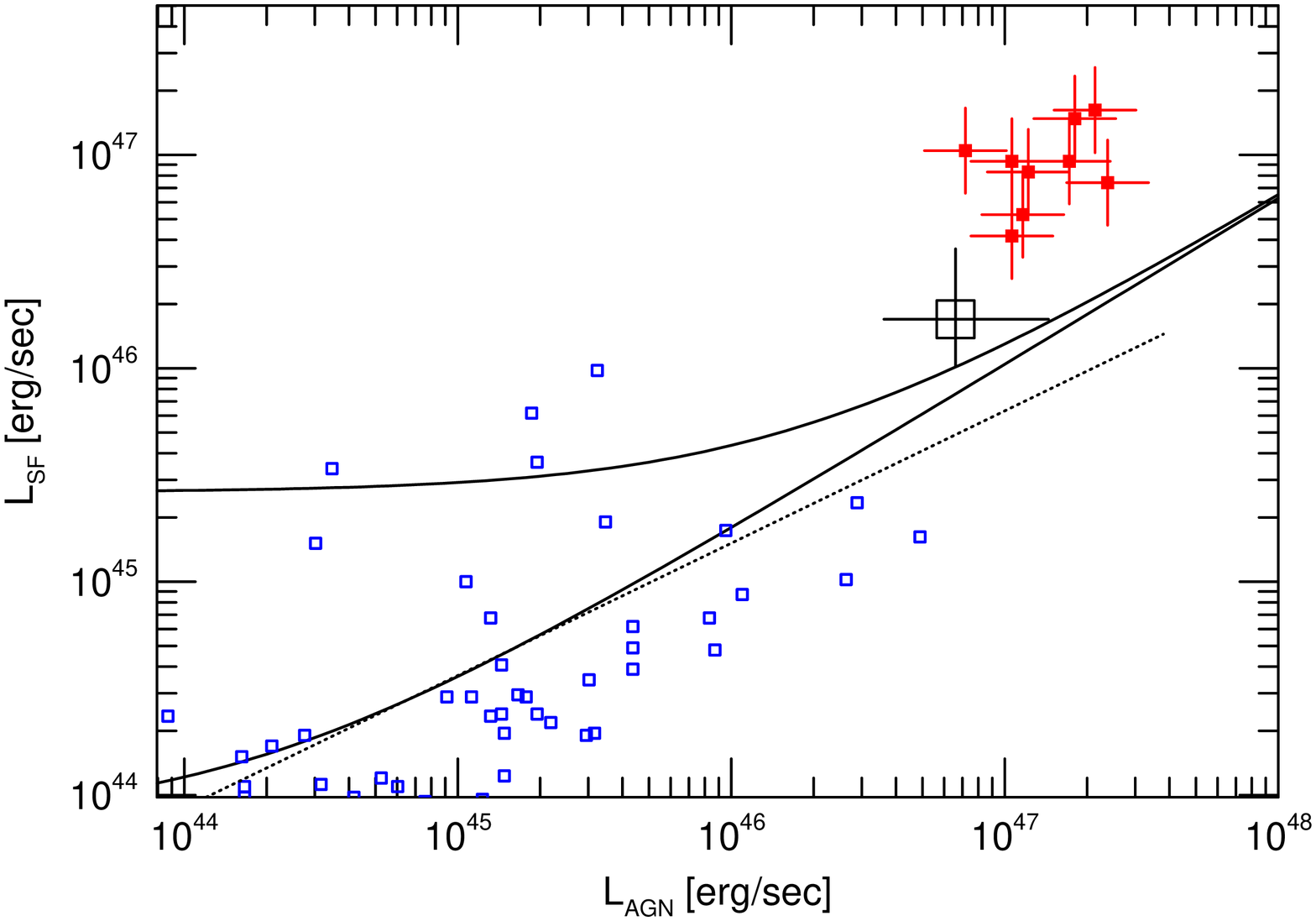}
\caption{
\LSF\ vs. \LAGN\ for 9 \herschel-detected sources at \zzz\ with measured \mbh\ (large red points), the mean of 26 \herschel-undetected sources
with measured \mbh\ (big black square), and 68 low redshift sources with known BH mass and \LAGN\ used in M12 where references
to these values are provided (blue squares). 
The solid curves are adopted from \cite{Rosario2012} and the redshift intervals, from bottom to top, are:
0--0.2, 0.2--0.5, 0.5--0.8, 0.8--1.5, and 1.5--2.5.
The dotted straight line is the \LAGN\ vs. \LSF\ relations for high-z sources from \cite{Wang2011}.
}
\label{fig:LSF_LAGN}
\end{figure}

 Fig.~\ref{fig:LSF_LAGN} shows the location of our sources in the \LSF-\LAGN\ plane. It also shows several curves
from  \cite{Rosario2012} for $z<0.5$
and $1.5 < z < 2.5$ sources\footnote{The curves were taken from \cite{Rosario2012} and scaled up by a factor 2 to allow for the
difference between L(60\mic) used in that paper and \LSF\ used here} and 68 AGNs with measured L(FIR) from the work of 
\cite{Mor2012a}, mostly local intermediate
luminosity objects. The local sources are quite similar in their properties to the brighter 
X-ray selected AGNs used in \cite{Rosario2012} but they do not represent a complete or flux limited sample and
are shown for illustration purpose only.
We also show a line based on \cite{Wang2011} which fits the most FIR luminous (as of 2011)
high-z sources.
This line is similar but somewhat shallower than the relationship suggested
by \cite{Netzer2009a}. We purposely avoid a comparison with very high-z sources observed at mm-wavelengths, e.g. the sample of \cite{Omont2013}, because of the unknown level of 
completeness of such samples.

To continue, we assume a simple conversion of mass accretion to \LAGN\ with efficiency $\eta_{BH}$ (the preferred value used here is 0.1).
We also assume the conversion of SFR to \LSF\ with efficiency $\eta_{SF}$ (equal to $7 \times 10^{-4}$ for the assumed
IMF).
Using this terminology we
can write the relative instantaneous growth rate of the stellar mass ($g(\mstar$)) and BH mass ($g(\mbh)$) in the following way:
\begin{equation}
\frac {g(\mstar)}{g(\mbh)} \simeq 140 
\left[ \frac{ 0.1(1-\eta_{BH})/\eta_{BH} } 
{7\times 10^{-4}(1-\eta_{SF})/\eta_{SF} } \right] 
\left[ \frac { \LSF}{\LAGN} \right] \, . 
\label{eq:relative_growth}
\end{equation}
As seen from Fig.~\ref{fig:LSF_LAGN}, the \herschel-detected \zzz\ sources are all around the location
where \LSF=\LAGN\ indicating $g(\mstar)/g(\mbh) \simeq 140$
(the mean values in Table~\ref{tab:most_likely} suggest $\sim 130$). The stacked source representing
about 75\% of our \zzz\ population is located at somewhat lower \LAGN\ and \LSF, very close to the correlation line of AGN dominated sources where 
$g(\mstar)/g(\mbh) \simeq 20$. Thus, unlike
earlier studies at lower redshift, \cite[e.g.][]{Rosario2012},
the most luminous AGNs at \zzz\ are hosted in galaxies with SF luminosity comparable to their AGN luminosity.

\begin{figure}[t]
\centering
\includegraphics[height=6cm]{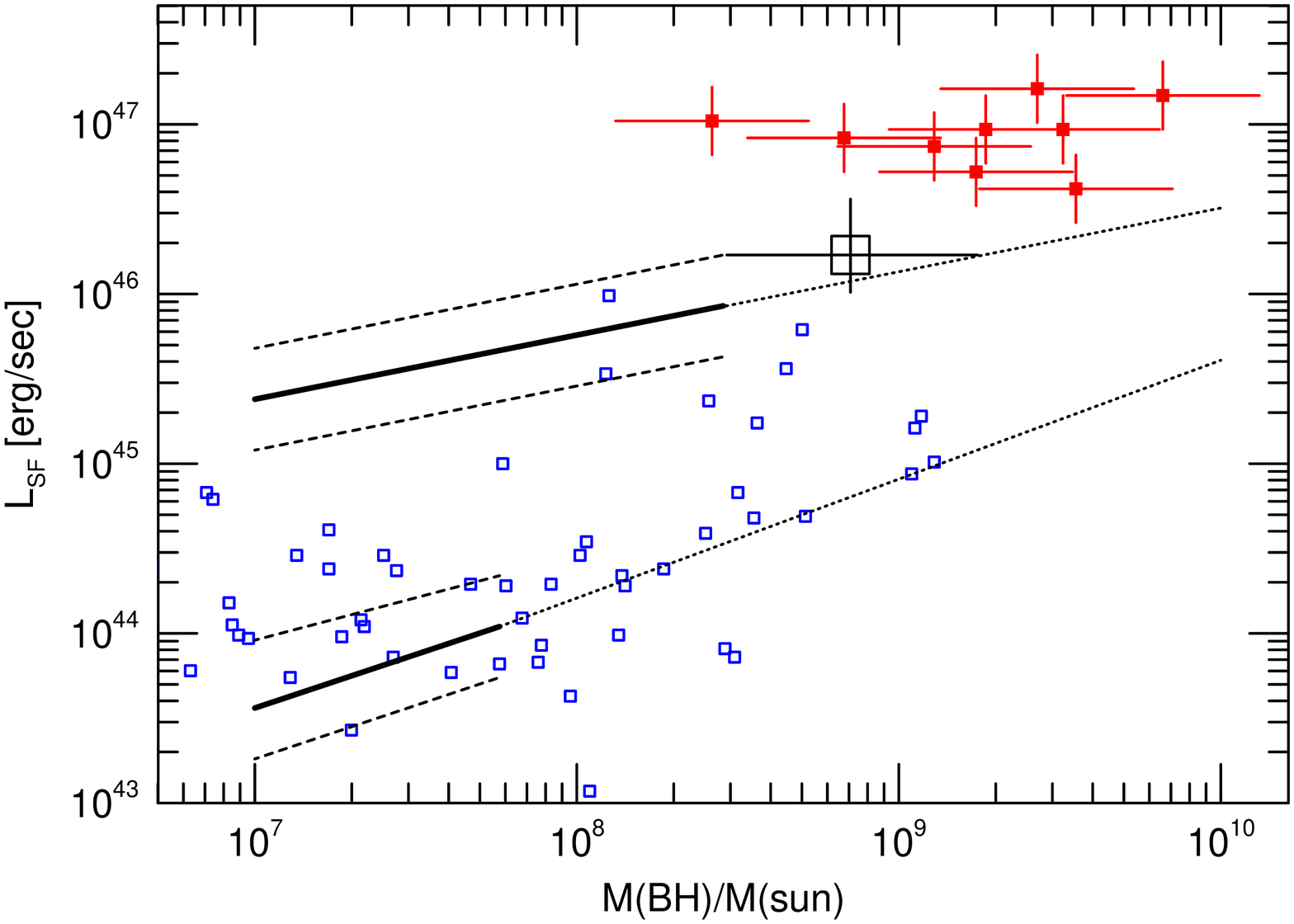}
\caption{Same samples and symbols as in Fig.~\ref{fig:LSF_LAGN} but for \LSF\ vs. \mbh.
The lower solid black line is obtained by assuming that all low redshifts AGNs are in MS SF galaxies where
\mstar/\mbh=700 (see text). The dashed lines show typical 0.3~dex uncertainties on the SFR for  MS objects.
 They extend to the largest
values of \mstar\ used in \cite{Whitaker2012}. The dotted line is a simple extrapolation of the solid line to larger masses. 
The upper straight lines are obtained by using the same assumptions but the expression for
the $z=2.5$ MS from \cite{Whitaker2012}.
 While the extrapolation (dotted line) 
covers the entire range of \mbh,
this is not very meaningful since galaxies of this mass (700 times the mass of the measured \mbh) have never been observed at high redshift.
}
\label{fig:LSF_MBH}
\end{figure}

We are also in a position to investigate the correlation between \mbh\ and \LSF\ at \zzz.
 This correlation is shown in Fig.~\ref{fig:LSF_MBH} and compared,
again, with the sample used by \cite{Mor2012a}. The tendency for larger active BHs to be associated with higher SFR is clear from the diagram. This is driven partly by
the correlation between \mbh\ and \LAGN\ and between \LAGN\ and \LSF\ of AGN dominated sources. 

Considering the low redshift sources,  
we can make the simplistic assumption that \mstar/\mbh=700, similar to the local relationship for bulge-dominated galaxies containing BHs with
masses in the range $10^7-10^8$\msun. We then used these derived values of \mstar, and the assumption  
\mstar$\sim M_{Bulge}$,  to compare with the known MS for SF galaxies at low redshift. The expression we used for the MS at
 $z=0$ is obtained from \cite{Whitaker2012}. This relationship is drawn as a straight solid line which
 goes roughly in the middle of the low redshift AGN distribution suggesting that  most of the hosts of these AGNs are MS SF galaxies.
The diagram also shows, in  dashed lines, the typical width of the MS obtained from \cite{Whitaker2012} and 
\cite{Rodighiero2011} (about 0.3~dex). We only draw these
lines up to the largest measured \mstar\ from \cite{Whitaker2012}, thus the largest mass part, reached by the dotted line,
is no more than simple extrapolation.
 
There is no simple way to estimate \mstar\ from \mbh\ for high redshift MS galaxies because the ratio of
\mstar/\mbh\ has never been directly measured beyond $z\sim 0.5$.
Nevertheless, for the sake of illustration only, we repeated the same exercise for $z=2.5$ using again the 
\cite{Whitaker2012} expressions. The results are shown as the upper straight solid line, with the same range of uncertainty on \LSF,
extending to the largest known \mstar\ from \cite{Whitaker2012}.
The dotted line is, again, a simple extrapolation of the solid line.
It falls close to the location of the undetected sources and well below the location of all detected sources.
This again is consistent with the idea that the detected sources are above the MS at that redshift.
We stress, again, that there is no observational justification based on direct measurement of \mbh, to this estimate of \mstar\ although some earlier
work \citep[e.g.][]{Aird2012} make such assumptions. Needless to say, there is no observational evidence for stellar masses as 
large as inferred from the dotted line ($5 \times 10^{12}$\msun) at such high redshifts.

\subsubsection{Cumulative BH and stellar mass growth}

Next we consider the evolution of the \zzz\ population to later times in the history of the universe.
For this we integrate $g(\mstar)$ and $g(\mbh)$ over cosmic and follow the \mstar/\mbh\ evolution 
using various estimated duty cycles for the two processes,
$\delta_{BH}$ and $\delta_{SF}$.
In particular, we are interested in the evolution between \zzz\ to \z24\ under the assumption, 
presented in T11, that the BHs studied here represent 
the population of the most massive BHs in the universe (some of which are not active). 
The goal is to test the suggestion that these BHs are hosted in the most massive galaxies and that 
beyond \z24, the BHs and the host galaxies
grow very slowly, or perhaps not at all. 

We experimented with 
both linear (constant \LAGN\ and \LSF) and exponential (constant \Ledd\ and sSFR) growth.
We illustrate the case that both BH groups grew their  mass by a factor 5 between \zzz\ and \z24\ and that 
throughout this period,  
$\eta_{BH}=0.1$. The calculations are similar to those discussed by T11 except for the fact that T11 considered individual BHs while here
we consider mean properties of two sub-groups. For the linear growth scenario we obtained the following numbers for the mean
BH duty cycles: $\delta_{BH} \sim 0.19$ 
for the  \herschel-detected sources and $\delta_{BH} \sim 0.15$ for the undetected sources. 
The corresponding duty cycles obtained for the exponential growth scenario are
$\delta_{BH} \simeq 0.095$ and $\delta_{BH} \simeq 0.085$, respectively.

The increase in stellar mass can be expressed as the stellar mass at time $t_2$ ($z=2.4$) relative to the one at time $t_1$ ($z=4.8$).
The growth can be linear (constant SFR) or exponential (constant sSFR) and we only illustrate the a case where
at
$t_1$ \mstar=$10^{11.5}$ \msun\ and at $t_2$ the mass is 5 times larger.
For exponential growth, we found the following corresponding time:
250 Myr for the \herschel-detected sources (duty cycle of about 0.17) and 1.16 Gyr for the undetected sources (duty cycle of about 0.80).
For linear growth we assume 
\begin{equation}
\mstar(t_2)=\int_{t_1}^{t_2} SFR(t-t_1)f_{loss}(t_2-t)dt \, ,
\label{eq:ms_growth}
\end{equation}
where $f_{loss}$ is a function that takes into account the stellar mass loss during the process.
This function decreases from 1 to about 0.5 over a period of $\sim 1 $Gyr for a \cite{Chabrier2003} IMF.
The calculated times are about 1.26 Gyr (duty cycle of about 0.85) for \herschel-detected sources and 5.7 Gyr 
(duty cycle greater than 1) for the undetected sources.
The latter is longer than the available time by 
a factor of almost 4 thus for these sources the largest stellar mass at $z=2.4$ is only about $6 \times 10^{11}$ \msun, i.e.,
below the mass of the most massive galaxies in the local universe.

The above numbers for the exponential growth scenarios indicate considerably shorter duty cycles for the BHs. 
This could in principle be verified
by an accurate census of active BHs and SF galaxies in such epochs. 
However, the predicted final point of exponential growth of the stellar mass
corresponds to enormous SFR of the host galaxies, way beyond what we measure at \zzz,
 sometimes between $z=4.8$ and $z=2.4$. We are not aware of any published work that present the observations
of such objects. It also involves the assumption
that most of this population is not forming stars at $z=2.4$ (i.e. they are quenched sources).
The linear growth combined with the assumption that we are observing MS host galaxies at \zzz\ is marginally consistent with SFR and \mstar\ measurements at $z \sim 2.4$ \citep{Wuyts2011a,Wuyts2011b}. 
This would mean that the most massive SF galaxies that are on the 
MS at those epochs are the hosts of the most massive BHs. In principle, this can be verified observationally. Here, again, a more complete census of SF galaxies vs. active
BHs will provide a constraint on the relative duty cycles of the two populations.

Finally, we comment on alternative scenarios to reach \mbh$\sim 10^{10}$ \msun\ that are different from the ones discussed here.
The BH population observed by us at \zzz\ may be at the end of their BH growth period and will never exceed by much the measured mass of $\sim 10^9$ \msun.
Another population of active BHs, that is not observed because they are totally obscured at \zzz, grows very rapidly and shed their ``cocoon'' at redshifts
2--3 where they are recognized to have \mbh$\sim 10^{10}$ \msun. Such a scenario has received much attention in various  theoretical studies 
\citep[e.g.][]{Hopkins2006a} but is
yet to be verified observationally.

\section{Summary and conclusions}

This paper reports the results of detailed \herschel/SPIRE observations of a unique flux limited
sample of 44 optically selected AGNs at \zzz\ known to contain the highest mass super massive BHs at this redshift.
 This is, so far, the largest narrow redshift range, high-$z$ sample observed by \herschel.
The observations are supplemented by our own \mbh\ measurements (T11), publicly available SDSS spectra, \wise\ photometry,
our own H-band and K-band spectroscopy, and newly obtained \spitzer/IRAC photometry. 
A subset of these observations was reported in M12 and the present paper completes the study of the sample.          

The results of our study are:
\begin{enumerate}
 \item 
Ten of the objects were detected by \herschel\ with \LSF\
in the range $10^{46.62}-10^{47.21}$ \ergs\ corresponding to SFR in the range 1090--4239 \msunyr.
Stacking analysis of 29 undetected sources results in significant signals in all the SPIRE bands.   
The corresponding mean \LSF\ is $10^{46.23}$ \ergs\ if no AGN contribution is assumed and
$10^{46.19}$\ergs\ if AGN contribution is taken into account.
We also tried T=40K gray-body fits to the 500\mic\ fluxes only. Such SEDs do not fit the other bands and their \LSF\ is
smaller by factors of $\sim 2$.
The remaining 5 sources show significant emission in the \herschel\ images that is not associated with the AGNs in the field.
\item
There is a clear correlation between BH mass, AGN luminosity and SFR in a sense that the mean \mbh\ and \LAGN\ of the 
\herschel-detected sources are higher than the ones of the undetected sources.
The luminosity differences between the detected and undetected sources 
are seen across the entire spectrum from rest-frame $\sim 900$\AA\ to FIR wavelengths.
\item 
The mean optical-UV SEDs of the two groups are 
similar to the spectrum of standard geometrically thin, optically thick accretion disks around BHs with the mean   
observed mass and accretion rates of the objects in those groups. 
\item
The transitions between MIR and FIR wavelengths in the mean SEDs of two groups are quite different.
The \herschel-detected objects show a more noticeable FIR bump and \LAGN$\simeq$ \LSF. We see indications of AGN-heated dust
in the short wavelength (rest-frame $\sim 43$\mic) bin of three of these sources.
The undetected sources have \LAGN$>$\LSF\ and a smoother, more gradual change from MIR to FIR wavelengths with possibly more
contribution by AGN-heated dust at 43\mic..
Unfortunately, the second result 
depends on an uncertain \wise\ photometry that was used to measure the MIR flux.
\item 
Out of various possible ways to explain the correlations of \LSF, \LAGN\ and \mbh, we chose to emphasize two that seem to agree
best with the properties of the \zzz\ sample. The first connects the
larger BH mass and AGN luminosity of the \herschel-detected sources to larger supply of cold gas to 
the entire galaxy and to the central source. This may be related to major mergers in these sources.
This scenario requires no feedback
to explain the differences between the two groups. The second explanation involves AGN feedback only in the 
hosts of the undetected sources. Because of the feedback, the 
sources are already beyond the peak of their SF and BH activity and hence the built-up of stellar and
BH mass are slower. However, the hypothetical feedback is not enough to totally quench SF and the overall properties of 
these host galaxies  are consistent with being on the SF MS at \zzz.  
\item
Our measured \mbh, \LAGN\ and \LSF, combined with the assumptions of little or no evolution of the MS 
at $z>2.5$, and mean \mstar=$10^{11-11.5}$\msun,
are consistent with the idea that the \herschel-detected sources 
represent SF galaxies that are above the MS at \zzz\ and most of the undetected sources are on the MS at this redshift.
Obviously, this is a rather speculative assumption given the lack of direct stellar mass measurements.
\item
Following the \zzz\ population to redshifts 2--3, assuming they end their evolution at those epochs 
by becoming the most massive BHs in the most massive (quenched) galaxies, lead to duty cycles that
are consistent with what is known about AGN and galaxy evolution. 
\end{enumerate}

\begin{acknowledgements}
This work is based on observations made with \herschel, 
a European Space Agency Cornerstone Mission with significant participation by NASA. 
Support for this work was provided by NASA through an award issued by JPL/Caltech. 
This work is also based on observations made with the 
\spitzer\ Space Telescope, which is operated by the Jet Propulsion Laboratory, 
California Institute of Technology under a contract with NASA. 
Support for this work was provided by NASA through an award issued by JPL/Caltech.
We are grateful to Raanan Nordon, Eyal Neistein, Dieter Lutz and David Rosario for useful discussions and comments. 
We thank an anonymous referee for useful suggestions that helped us to improve the paper.
We thank the DFG for support via German Israeli Cooperation grant STE1869/1-1.GE625/15-1. 
Funding for this work has also been provided by the Israel Science Foundation grants 364/07 and 284/13.
BT acknowledges support by the Benoziyo center for Astrophysics.
\end{acknowledgements}

\end{document}